\newif\ifemulate
\emulatetrue

\ifemulate
   \documentclass{emulateapj}
   \slugcomment{Accepted for Publication in ApJ}
\else
   \documentclass[12pt, preprint]{aastex}
\fi

\lefthead{Rudnick et al.}
\righthead{Cosmic Mass Density Evolution}

\usepackage{graphicx, xspace, epsf}

\newcommand\lrest{${\rm L^{rest}_\lambda}$\xspace}

\renewcommand{\l}[1]{${\rm L}_{\rm #1}^{rest}$\xspace}

\newcommand\bvr{$(B~-~V)_{rest}$\xspace}
\newcommand\ubr{$(U~-~B)_{rest}$\xspace}

\newcommand\uvr{$(U~-~V)_{rest}$\xspace}
\newcommand\zp{$z_{phot}$\xspace}
\newcommand\dzp{$\delta z_{phot}$\xspace}

\newcommand\Lya{${\rm Ly\alpha}$\xspace}
\newcommand\etal {et al.\xspace}
\newcommand\Js{$J_s$\xspace}
\newcommand\Jsab{$J_{s,{\rm AB}}$\xspace}
\newcommand\Hab{$H_{{\rm AB}}$\xspace}
\newcommand\Ks{$K_{s}$\xspace}
\newcommand\Ksab{$K_{s,{\rm AB}}$\xspace}
\newcommand\Ksvega{$K_{s,vega}$\xspace}
\newcommand\Hvega{$H_{vega}$\xspace}
\newcommand\Jsvega{$J_{s,vega}$\xspace}
\newcommand\Ksabtot{$K_{s,{\rm AB}}^{\rm tot}$\xspace}

\newcommand\ang{${\rm \AA}$\xspace}
\newcommand\zs{$z_{spec}$\xspace}

\newcommand\lstar{${\rm L}_*$\xspace}
\newcommand\lstarlocal{${\rm L}_*^{\rm local}$\xspace}
\newcommand{\lstarlocallam}[1]{${\rm L}_{*,#1}^{\rm local}$\xspace}
\newcommand{\lstarsdss}[1]{${\rm L}_{*,#1}^{\rm SDSS}$\xspace}
\newcommand{\lstarlam}[1]{${\rm L}_{*,#1}$\xspace}

\newcommand\mlstar{${\rm M}/{\rm L}^*$\xspace}
\newcommand{\mlstarlam}[1]{${\rm M}/{\rm L}^*_{#1}$\xspace}

\newcommand\chisq{$\chi^2$\xspace}

\newcommand\eg{e.g.,\xspace}

\newcommand\uv{$(U~-~V)$\xspace}
\newcommand\lumsol{${\rm ~h_{70}^{-2}~L_{\odot}}$\xspace}
\newcommand\rhostar{$\rho_*$\xspace}

\newcommand\jrest{$j^{\rm rest}_\lambda$\xspace}
\newcommand{\jrestlam}[1]{$j^{\rm rest}_{\rm #1}$\xspace}
\newcommand{\jsdss}[1]{$j_{#1}^{\rm SDSS}$\xspace}
\newcommand\magu{$U_{300}$\xspace}
\newcommand\magb{$B_{450}$\xspace}
\newcommand\magv{$V_{606}$\xspace}
\newcommand\magi{$I_{814}$\xspace}
\renewcommand\u{$U$\xspace}
\newcommand\mb{$B$\xspace}

\newcommand\mv{$V$\xspace}

\newcommand\mi{$I$\xspace}

\newcommand\mh{$H$\xspace}

\newcommand\lthresh{${\rm L}_{\lambda}^{\rm thresh}$\xspace}
\newcommand{\lthreshlam}[1]{${\rm L}_{#1}^{\rm thresh}$\xspace}

\begin{document}
\title{The Rest-Frame Optical Luminosity Density, Color, and Stellar Mass Density of the Universe from z=0 to z=3 \altaffilmark{1}}

\author{Gregory Rudnick\altaffilmark{2}, 
Hans-Walter Rix\altaffilmark{3}, 
Marijn Franx\altaffilmark{4}, 
Ivo Labb\'{e}\altaffilmark{4},  
Michael Blanton\altaffilmark{5}, 
Emanuele Daddi\altaffilmark{6},
Natascha M. F\"orster Schreiber\altaffilmark{4}, 
Alan Moorwood\footnotemark[6], 
Huub R\"ottgering\altaffilmark{4}, 
Ignacio Trujillo\altaffilmark{3},
Arjen van de Wel\altaffilmark{4}, 
Paul van der Werf\altaffilmark{4}, 
Pieter G. van Dokkum\altaffilmark{7}, 
\& Lottie van Starkenburg\altaffilmark{4}}

\altaffiltext{1}{Based on observations with the
    NASA/ESA \textit{Hubble Space Telescope}, obtained at the Space
    Telescope Science Institute, which is operated by the AURA, Inc.,
    under NASA contract NAS5-26555.  Also based on observations
    collected at the European Southern Observatories on Paranal, Chile
    as part of the ESO programme 164.O-0612}

\altaffiltext{2}{Max-Planck-Institut f\"{u}r Astrophysik, Karl-Schwarzschild-Strasse 1, Garching, D-85741, Germany, \texttt{grudnick@mpa-garching.mpg.de}}
\altaffiltext{3}{Max-Planck-Institut f\"{u}r Astronomie, K\"{o}nigstuhl 17, Heidelberg, D-69117, Germany}
\altaffiltext{4}{Leiden Observatory, PO BOX 9513, 2300 RA Leiden, Netherlands}
\altaffiltext{5}{Department of Physics, New York University, 4 Washington Place, New York, NY 10003}
\altaffiltext{6}{European Southern Observatory, Karl-Schwarzschild-Strasse 2, 85748 Garching, Germany}
\altaffiltext{7}{Astronomy Department, Yale University, P.O. Box 208101, New Haven, CT 06520-8101}

\begin{abstract}

 We present the evolution of the rest-frame optical luminosity
 density, \jrest, of the integrated rest-frame optical color, and of
 the stellar mass density, \rhostar, for a sample of \Ks-band selected
 galaxies in the HDF-S.  We derived \jrest in the rest-frame \u, \mb,
 and \mv-bands and found that \jrest increases by a factor of
 $1.9\pm0.4$, $2.9\pm0.6$, and $4.9\pm1.0$ in the \mv, \mb, and \u
 rest-frame bands respectively between a redshift of 0.1 and 3.2.  We
 derived the luminosity weighted mean cosmic \ubr and \bvr colors as a
 function of redshift.  The colors bluen almost monotonically with
 increasing redshift; at $z=0.1$, the \ubr and \bvr colors are 0.16
 and 0.75 respectively, while at $z=2.8$ they are -0.39 and 0.29
 respectively.  We derived the luminosity weighted mean \mlstarlam{V}
 using the correlation between \uvr and log$_{10}$\mlstarlam{V} which
 exists for a range in smooth SFHs and moderate extinctions.  We have
 shown that the mean of individual \mlstarlam{V} estimates can
 overpredict the true value by $\sim 70\%$ while our method
 overpredicts the true values by only $\sim 35\%$.  We find that the
 universe at $z\sim3$ had $\sim 10$ times lower stellar mass density
 than it does today in galaxies with \l{V}$>1.4 \times
 10^{10}$\lumsol.  50\% of the stellar mass of the universe was formed
 by $z\sim 1-1.5$.  The rate of increase in \rhostar with decreasing
 redshift is similar to but above that for independent estimates from
 the HDF-N, but is slightly less than that predicted by the integral
 of the SFR($z$) curve.
\end{abstract}
\section{Introduction}
\label{Intro}

 A primary goal of galaxy evolution studies is to elucidate how the
 stellar content of the present universe was assembled over time.
 Enormous progress has been made in this field over the past decade,
 driven by advances over three different redshift ranges.  Large scale
 redshift surveys with median redshifts of $z\sim 0.1$ such as the
 Sloan Digital Sky Survey (SDSS; York et al. 2000) and the 2dF Galaxy
 Redshift Survey (2dFGRS; Colless et al. 2001), coupled with the near
 infrared (NIR) photometry from the 2 Micron All Sky Survey (2MASS;
 Skrutskie et al. 1997), have recently been able to assemble the
 complete samples, with significant co-moving volumes, necessary to
 establish crucial local reference points for the local luminosity
 function (e.g. Folkes et al. 1999; Blanton et al. 2001; Norberg et
 al. 2002; Blanton et al. 2003c) and the local stellar mass function of
 galaxies \citep{Cole01, Bell03a}.

 At $z\lesssim 1$, the pioneering study of galaxy evolution was the
 Canada France Redshift Survey (CFRS; Lilly et al. 1996).  The
 strength of this survey lay not only in the large numbers of galaxies
 with confirmed spectroscopic redshifts, but also in the \mi-band
 selection, which enabled galaxies at $z\lesssim 1$ to be selected in
 the rest-frame optical, the same way in which galaxies are selected
 in the local universe.

 At high redshifts the field was revolutionized by the identification,
 and subsequent detailed follow-up, of a large population of
 star-forming galaxies at $z>2$ \citep{Stei96}.  These Lyman Break
 Galaxies (LBGs) are identified by the signature of the redshifted
 break in the far UV continuum caused by intervening and intrinsic
 neutral hydrogen absorption.  There are over 1000 spectroscopically
 confirmed LBGs at $z>2$, together with the analogous U-dropout
 galaxies identified using Hubble Space Telescope (HST) filters.  The
 individual properties of LBGs have been studied in great detail.
 Estimates for their star formation rates (SFRs), extinctions, ages,
 and stellar masses have been estimated by modeling the broad band
 fluxes (Sawicki \& Yee 1998; hereafter SY98; Papovich et al. 2001;
 hereafter P01; Shapley et al. 2001).  Independent measures of their
 kinematic masses, metallicities, SFRs, and initial mass functions
 (IMFs) have been determined using rest-frame UV and optical
 spectroscopy \citep{Pett00,Shap01,Pett01,Pett02,Erb03}.

 Despite these advances, it has proven difficult to reconcile the
 ages, SFRs, and stellar masses of individual galaxies at different
 redshifts within a single galaxy formation scenario.  Low redshift
 studies of the fundamental plane indicate that the stars in
 elliptical galaxies must have been formed by $z>2$ (\eg van Dokkum
 \etal 2001) and observations of evolved galaxies at $1<z<2$ indicate
 that the present population of elliptical galaxies was already in
 place at $z\gtrsim2.5$ (\eg Ben\'{i}tez \etal 1999; Cimatti et
 al. 2002; but see Zepf 1997).  In contrast, studies of star forming
 Lyman Break Galaxies spectroscopically confirmed to lie at $z>2$
 (LBGs; Steidel \etal 1996, 1999) claim that LBGs are uniformly very
 young and a factor of 10 less massive than present day \lstar
 galaxies (\eg SY98; P01; Shapley \etal 2001).

 An alternative method of tracking the build-up of the cosmic stellar
 mass is to measure the total emissivity of all relatively unobscured
 stars in the universe, thus effectively making a luminosity weighted
 mean of the galaxy population.  This can be partly accomplished by
 measuring the evolution in the global luminosity density $j(z)$ from
 galaxy redshift surveys.  Early studies at intermediate redshift have
 shown that the rest-frame UV and \mb-band $j(z)$ are steeply
 increasing out to $z\sim1$ (\eg Lilly \etal 1996; Fried \etal 2001).
 \citet{Wolf03} has recently measured $j(z)$ at $0<z<1.2$ from the
 COMBO-17 survey using $\sim $25,000 galaxies with redshifts accurate
 to $\sim 0.03$ and a total area of 0.78 degrees.  At rest-frame
 2800\ang these measurements confirm those of \citet{Lill96} but do
 not support claims for a shallower increase with redshift which goes
 like $(1+z)^{1.5}$ as claimed by \citet{Cow99} and \citet{Wil02}.  On
 the other hand, the \mb-band evolution from \citet{Wolf03} is only a
 factor of $\sim 1.6$ between $0<z<1$, considerably shallower than the
 factor of $\sim 3.75$ increase seen by \citet{Lill96}.  At $z>2$
 measurements of the rest-frame UV $j(z)$ have been made using the
 optically selected LBG samples (\eg Madau \etal 1996; Sawicki, Lin,
 \& Yee 1997; Steidel \etal 1999; Poli \etal 2001) and NIR selected
 samples \citep{kashikawa03,Poli03,Thom03} and, with modest extinction
 corrections, the most recent estimates generically yield rest-frame
 UV $j(z)$ curves which, at $z>2$, are approximately flat out to
 $z\sim 6$ (cf. Lanzetta \etal 2002).  Dickinson \etal (2003;
 hereafter D03) have used deep NIR data from NICMOS in the HDF-N to
 measure the rest-frame \mb-band luminosity density out to $z\sim 3$,
 finding that it remained constant to within a factor of $\sim 3$.  By
 combining $j(z)$ measurements at different rest-frame wavelengths and
 redshifts, \citet{Mad98} and \citet{Pei99} modeled the emission in
 all bands using an assumed global SFH and used it to constrain the
 mean extinction, metallicity, and IMF.  \citet{Bolzonella02} measured
 NIR luminosity functions in the HDF-N and HDF-S and find little
 evolution in the bright end of the galaxy population and no decline
 in the rest-frame NIR luminosity density out to $z\sim2$.  In
 addition, \citet{Baldry02} and \citet{Gla03a} have used the mean
 optical cosmic spectrum at $z\sim 0$ from the 2dFGRS and the SDSS
 respectively to constrain the cosmic star formation history.

 Despite the wealth of information obtained from studies of the
 integrated galaxy population, there are major difficulties in using
 these many disparate measurements to re-construct the evolution in
 the stellar mass density.  First, and perhaps most important, the
 selection criteria for the low and high redshift surveys are usually
 vastly different.  At $z<1$ galaxies are selected by their rest-frame
 optical light.  At $z>2$, however, the dearth of deep, wide-field NIR
 imaging has forced galaxy selection by the rest-frame UV light.
 Observations in the rest-frame UV are much more sensitive to the
 presence of young stars and extinction than observations in the
 rest-frame optical.  Second, state-of-the-art deep surveys have only
 been performed in small fields and the effects of field-to-field
 variance at faint magnitudes, and in the rest-frame optical, are not
 well understood.

 In the face of field-to-field variance, the globally averaged
 rest-frame color may be a more robust characterization of the galaxy
 population than either the luminosity density or the mass density
 because it is, to the first order, insensitive to the exact density
 normalization.  At the same time, it encodes information about the
 dust obscuration, metallicity, and SFH of the cosmic stellar
 population.  It therefore provides an important constraint on galaxy
 formation models which may be reliably determined from relatively
 small fields.

 To track consistently the globally averaged evolution of the galaxies
 which dominate the stellar mass budget of the universe -- as opposed
 to the UV luminosity budget -- over a large redshift range a
 different strategy than UV selection must be adopted.  It is not only
 desirable to measure $j(z)$ in a constant rest-frame optical
 bandpass, but it is also necessary that galaxies be selected by light
 redward of the Balmer/4000\ang break, where the light from older stars
 contributes significantly to the SED.  To accomplish this, we
 obtained ultra-deep NIR imaging of the WFPC2 field of the HDF-S
 \citep{Cas00} with the Infrared Spectrograph And Array Camera (ISAAC;
 Moorwood \etal 1997) at the Very Large Telescope (VLT) as part of the
 Faint Infrared Extragalactic Survey (FIRES; Franx \etal 2000).  The
 FIRES data on the HDF-S, detailed in Labb\'e \etal (2003; hereafter
 L03), provide us with the deepest ground-based \Js and \mh data and
 the overall deepest \Ks-band data in any field allowing us to reach
 rest-frame optical luminosities in the \mv-band of $\sim
 0.6$~\lstarlocal at $z\sim 3$.  First results using a smaller set of
 the data were presented in Rudnick \etal (2001; hereafter R01).  The
 second FIRES field, centered on the $z=0.83$ cluster MS1054-03, has
 $\sim 1$ magnitude less depth but $\sim5$ times greater area
 \citep{Foerster03}.

 In the present work we will draw on photometric redshift estimates,
 \zp for the \Ks-band selected sample in the HDF-S (R01; L03), and on
 the observed SEDs, to derived rest-frame optical luminosities \lrest
 for a sample of galaxies selected by light redder than the rest-frame
 optical out to $z\sim3$.  In \S~\ref{data} we describe the
 observations, data reduction, and the construction of a \Ks-band
 selected catalog with $0.3-2.2\mu m$ photometry, which selects
 galaxies at $z<4$ by light redward of the 4000\ang break.  In
 \S~\ref{methodsec} we describe our photometric redshift technique,
 how we estimate the associated uncertainties in \zp, and how we
 measure \lrest for our galaxies.  In \S~\ref{results} we use our
 measures of \lrest for the individual galaxies to derive the mean
 cosmic luminosity density, \jrest and the cosmic color and then use
 these to measure the stellar mass density \rhostar as a function of
 cosmic time.  We discuss our results in \S~\ref{discuss} and
 summarize in \S~\ref{summary}.  Throughout this paper we assume
 $\Omega_\mathrm{M}=0.3,~\Omega_{\Lambda}=0.7,~\mathrm{and~H_o}=70~{\rm
 h_{70}~km~s^{-1} Mpc^{-1}}$ unless explicitly stated otherwise.

\section{Data}
\label{data}

 A complete description of the FIRES observations, reduction
 procedures, and the construction of photometric catalogs is presented
 in detail in L03; we outline the important steps below.

 Objects were detected in the \Ks-band image with version 2.2.2 of the
 SExtractor software \citep{Ber96}.  For consistent photometry between
 the space and ground-based data, all images were then convolved to
 $0\farcs48$, the seeing in our worst NIR band.  Photometry was then
 performed in the \magu, \magb, \magv, \magi, \Js, \mh, and \Ks-band
 images using specially tailored isophotal apertures defined from the
 detection image.  In addition, a measurement of the total flux in the
 \Ks-band, \Ksabtot, was obtained using an aperture based on the
 SExtractor \textit{AUTO} aperture\footnote{The reduced images,
 photometric catalogs, photometric redshift estimates, and rest-frame
 luminosities are available online through the FIRES homepage at
 \texttt{http://www.strw.leidenuniv.nl/$\sim$fires}.}.  Our effective area
 is 4.74 square arcminutes, including only areas of the chip which
 were well exposed.  All magnitudes are quoted in the Vega system
 unless specifically noted otherwise.  Our adopted conversions from
 Vega system to the AB system are \Jsvega~= \Jsab~- 0.90, \Hvega~=
 \Hab~- 1.38, and \Ksvega~= \Ksab~- 1.86 \citep{BB88}.

\section{Measuring Photometric Redshifts and Rest-Frame Luminosities}
\label{methodsec}

\subsection{Photometric Redshift Technique}
\label{zpmethod}

 We estimated \zp from the broad-band SED using the method described
 in R01, which attempts to fit the observed SED with a linear
 combination of redshifted galaxy templates.  We made two
 modifications to the R01 method.  First, we added an additional
 template constructed from a 10 Myr old, single age, solar metallicity
 population with a \citet{Sal55} initial mass function (IMF) based on
 empirical stellar spectra from the 1999 version of the \citet{BC93}
 stellar population synthesis code.  Second, a 5\% minimum flux error
 was adopted for all bands to account for the night-to-night
 uncertainty in the derived zeropoints and for template mismatch
 effects, although in reality both of these errors are non-gaussian.

 Using 39 galaxies with reliable FIRES photometry and spectroscopy
 available from \citet{Cris00}, \citet{Rig00},
 \citet{Gla03b}\footnote{available at
 \texttt{http://www.aao.gov.au/hdfs/}}, \citet{Van02}, and
 \citet{Rud03} we measured the redshift accuracy \ of our technique to
 be $\left\langle~|z_{spec} -
 z_{phot}|~/~(1+z_{spec})~\right\rangle=0.09$ for $z<3$ .  There is
 one galaxy at \zs=2.025 with \zp$=0.12$ but with a very large
 internal \zp uncertainty.  When this object is removed,
 $\left\langle~|z_{spec} -
 z_{phot}|~/~(1+z_{spec})~\right\rangle=0.05$ at \zs$>1.3$.

 For a given galaxy, the photometric redshift probability distribution
 can be highly non-Gaussian and contain multiple \chisq minima at
 vastly different redshifts.  An accurate estimate of the error in \zp
 must therefore not only contain the two-sided confidence interval in
 the local \chisq minimum, but also reflect the presence of alternate
 redshift solutions.  The difficulties of measuring the uncertainty in
 \zp were discussed in R01 and will not be repeated in detail here.
 To improve on R01, however, we have developed a Monte Carlo method
 which takes into account, on a galaxy-by-galaxy basis, flux errors
 and template mismatch.  These uncertainty estimates are called \dzp.
 For a full discussion of this method see Appendix~\ref{app_z_err}.

 Galaxies with \Ksabtot$\geq 25$ have such high photometric errors
 that the \zp estimates can be very uncertain.  At \Ksabtot$<25$,
 however, objects are detected at better than the 10-sigma level and
 have well measured NIR SEDs, important for locating redshifted
 optical breaks.  For this reason, we limited our catalog to the 329
 objects that have \Ksabtot$<25$, lie on well exposed sections of the
 chip, and are not identified as stars (see \S\ref{starid}).

\subsubsection{Star Identification}
\label{starid}

 To identify probable stars in our catalog we did not use the profiles
 measured from the WFPC2-imaging because it is difficult to determine
 the size at faint levels.  At the same time, we verified that the
 stellar template fitting technique identified all bright unsaturated
 stars in the image.  Instead, we compared the observed SEDs with
 those from 135 NextGen version 5.0 stellar atmosphere models
 described in \citet{Haus99} and available at
 \texttt{http://dilbert.physast.uga.edu/$\sim$yeti/mdwarfs.html}.  We
 used models with log$(g)$ of 5.5 and 6, effective temperatures
 ranging from 1600 K to 10,000 K, and metallicities of solar and
 1/10th solar.  We identified an object as a stellar candidate if the
 raw \chisq of the stellar fit was lower than that of the best-fit
 galaxy template combination.  Four of the stellar candidates from
 this technique (objects 155, 230, 296, and 323) are obviously
 extended and were excluded from the list of stellar candidates.  Two
 bright stars (objects 39 and 51) were not not identified by this
 technique because they are saturated in the HST images and were added
 to the list by hand.  We ended up with a list of 29 stars that had
 \Ksabtot$<25$ and lie on well exposed sections of the chip.  These
 were excluded from all further analysis.

\subsection{Rest-Frame Luminosities}
\label{restlum}

 To measure the \lrest of a galaxy one must combine its redshift with
 the observed SED to estimate the intrinsic SED.  In practice, this
 requires some assumptions about the intrinsic SED.

 In R01 we derived rest-frame luminosities from the best-fit
 combination of spectral templates at \zp, which assumes that the
 intrinsic SED is well modeled by our template set.  We know that for
 many galaxies the best-fit template matches the position and strength
 of the spectral breaks and the general shape of the SED.  There are,
 however, galaxies in our sample which show clear residuals from the
 best fit template combination.  Even for the qualitatively good fits,
 the model and observed flux points can differ by $\sim 10\%$,
 corresponding to a $\sim 15\%$ error in the derived rest-frame color.
 As we will see in \S\ref{color}, such color errors can cause errors
 of up to a factor of 1.5 in the \mv-band stellar mass-to-light ratio,
 \mlstarlam{V}.

 Here we used a method of estimating \lrest which does not depend
 directly on template fits to the data but, rather, interpolates
 directly between the observed bands using the templates as a guide.
 We define our rest-frame photometric system in Appendix~\ref{photsys}
 and explain our method for estimating \lrest in
 Appendix~\ref{lumder}.

 We plot in Figure~\ref{lumvol} the rest-frame luminosities vs.
 redshift and enclosed volume for the \Ksabtot~$<25$ galaxies in the
 FIRES sample.  The different symbols represent different \dzp values
 and since the derived luminosity is tightly coupled to the redshift,
 we do not independently plot \lrest errorbars.  The tracks indicate
 the \lrest for different SED types normalized to \Ksabtot$=25$, while
 the intersection of the tracks in each panel indicates the redshift
 at which the rest-frame filter passes through our \Ks-band detection
 filter.  There is a wide range in \lrest at all redshifts and there
 are galaxies at $z>2$ with \lrest much in excess of the local \lstar
 values.  Using the full FIRES dataset, we are much more sensitive
 than in R01; objects at $z\approx 3$ with \Ksabtot~$=25$ have
 \l{V}$\approx 0.6 * $\lstarlocallam{V}, as defined from the z=0.1
 sample of Blanton et al. (2003c; hereafter B03).  As seen in R01
 there are many galaxies at $z>2$, in all bands, with
 \lrest$\geq$\lstarlocal.  R01 found 10 galaxies at $2\leq z\leq 3.5$
 with \l{B}$> 10^{11}$\lumsol and inferred a brightening in the
 luminosity function of $\sim 1-1.3$ magnitudes.  We confirm their
 result when using the same local luminosity function \citep{blan01}.
 Although this brightening is biased upwards by photometric redshift
 errors, we find a similar brightening of approximately $\sim 1$
 magnitude after correction for this effect.  As also noticed in R01,
 we found a deficit of luminous galaxies at $1.5 \lesssim z \lesssim
 2$ although this deficit is not as pronounced at lower values of
 \lrest.  The photometric redshifts in the HDF-S, however, are not
 well tested in this regime.  To help judge the reality of this
 deficit we compared our photometric redshifts on an object-by-object
 basis to those of the Rome group \citep{Fon00}\footnote{available at
 \texttt{http://www.mporzio.astro.it/HIGHZ/HDF.html}} who derived \zp
 estimates for galaxies in the HDF-S using much shallower NIR data.
 We find generally good agreement in the \zp estimates, although there
 is a large scatter at $1.5<z<2.0$.  Both sets of photometric
 redshifts show a deficit in the \zp distribution, although the Rome
 group's gap is less pronounced than ours and is at a slightly lower
 redshift.  In addition, we examined the photometric redshift
 distribution of the NIR selected galaxies of D03 in the HDF-N, which
 have very deep NIR data.  These galaxies also showed a gap in the \zp
 distribution at $z\sim1.6$.  Together these results indicate that
 systematic effects in the \zp determinations may be significant at
 $1.5<z<2.0$.  On the other hand, we also derived photometric
 redshifts for a preliminary set of data in the MS1054-03 field of the
 FIRES survey, whose filter set is similar, but which has a \u instead
 of \magu filter.  In this field, no systematic depletion of $1.5<z<2$
 galaxies was found.  It is therefore not clear what role systematic
 effects play in comparison to field-to-field variations in the true
 redshift distribution over this redshift range.  Obtaining
 spectroscopic redshifts at $1.5<z<2$ is the only way to judge the
 accuracy of the \zp estimates in this regime.

 We have also split the points up according to whether or not they
 satisfied the U-dropout criteria of \citet{Gia01} which were designed
 to pick unobscured star-forming galaxies at $z\gtrsim 2$.  As
 expected from the high efficiency of the U-dropout technique, we find
 that only 15\% of the 57 classified U-dropouts have \zp$<2$.  As we
 will discuss in \S\ref{lumdens} we measured the luminosity density
 for objects with \l{V}$>1.4 \times 10^{10}$\lumsol.  Above this
 threshold, there are 62 galaxies with $2<z<3.2$, of which 26 are not
 classified as U-dropouts.  These non U-dropouts number among the most
 rest-frame optically luminous galaxies in our sample.  In fact, the
 most rest-frame optically luminous object at $z<3.2$ (object 611) is
 a galaxy which fails the U-dropout criteria.  10 of these 26 objects,
 including object 611, also have $J-K>2.3$, a color threshold which
 has been shown by \citet{Franx03} and \citet{dokkum03} to efficiently
 select galaxies at $z>2$.  These galaxies are not only luminous but
 also have red rest-frame optical colors, implying high \mlstar
 values.  \citet{Franx03} showed that they likely contribute
 significantly ($\sim 43\%$) to the stellar mass budget at high
 redshifts.

\subsubsection{Emission Lines}
\label{emis}

 There will be emission line contamination of the rest-frame
 broad-band luminosities when rest-frame optical emission lines
 contribute significantly to the flux in our observed filters.  P01
 estimated the effect of emission lines in the NICMOS F160W filter and
 the \Ks filter and found that redshifted, rest-frame optical emission
 lines, whose equivalent widths are at the maximum end of those
 observed for starburst galaxies (rest-frame equivalent width $\sim
 200$\ang), can contribute up to 0.2 magnitudes in the NIR filters.
 In addition, models of emission lines from \citet{Charlot01} show
 that emission lines will tend to drive the \ubr color to the blue
 more easily than the \bvr color for a large range of models.  Using
 the $UBV$ photometry and spectra of nearby galaxies from the Nearby
 Field Galaxy Survey (NFGS; Jansen et al. 2000a; Jansen et al. 2000b)
 we computed the actual correction to the \ubr and \bvr colors as a
 function of \bvr.  For the bluest galaxies in \bvr, emission lines
 bluen the \ubr colors by $\sim 0.05$ and the \bvr colors only by
 $<0.01$.  Without knowing beforehand the strength of emission lines
 in any of our galaxies, we corrected our rest-frame colors based on
 the results from Jansen et al.  We ignored the very small correction
 to the \bvr colors and corrected the \ubr colors using the equation:
\begin{equation}
 (U-B)_{corrected} = (U-B) - 0.0658 \times (B-V) + 0.0656
\label{emiseq}
\end{equation}
 which corresponds to a linear fit to the NFGS data.  These effects
 might be greater for objects with strong AGN contribution to their
 fluxes.

\begin{figure}
\epsscale{.80}
\plotone{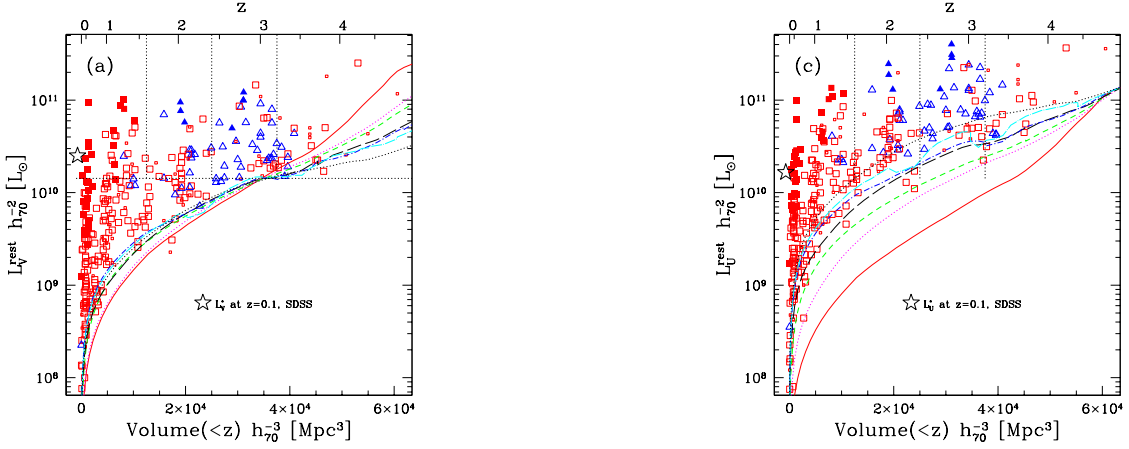}
\caption {The distribution of rest-frame \mv, \mb, and \u-band
 luminosities as a function of enclosed co-moving volume and \zp is
 shown in figures (a), (b), and (c) respectively for galaxies with
 \Ksabtot$<25$.  Galaxies which have spectroscopic redshifts are
 represented by solid points and for these objects \lrest is measured
 at \zs.  Large symbols have $\delta z_{phot}/(1 + z_{phot})<0.16$ and
 small symbols have $\delta z_{phot}/(1 + z_{phot})\geq0.16$.
 Triangle points would be classified as U-dropouts according to the
 selection of \citet{Gia01}.  As is expected, most of the galaxies
 selected as U-dropouts have \zp$\gtrsim2$.  Note, however, the large
 numbers of rest-frame optically luminous galaxies at $z>2$ which
 would not be selected as U-dropouts.  The large stars in each panel
 indicate the value of \lstarlocal from \citet{blanton03c}.  In the
 \mv-band we are sensitive to galaxies at 60\% of \lstarlocal even at
 $z\sim3$ and there are galaxies at \zp$\geq 2$ with \lrest $\geq
 10^{11}$\lumsol.  The tracks represent the values of \lrest for our
 seven template spectra normalized at each redshift to \Ksabtot$=25$.
 The specific tracks correspond to the E (solid), Sbc (dot), Scd
 (short dash), Im (long dash), SB1 (dot--short dash), SB2 (dot--long
 dash), and 10my (dot) templates.  The horizontal dotted line in (a)
 indicates the luminosity threshold \lthreshlam{V} above which we
 measure the rest-frame luminosity density \jrest and the vertical
 dotted lines in each panel mark the redshift boundaries of the
 regions for which we measure \jrest.  }
\label{lumvol}
\end{figure}

\begin{figure}
\epsscale{.80}
\plotone{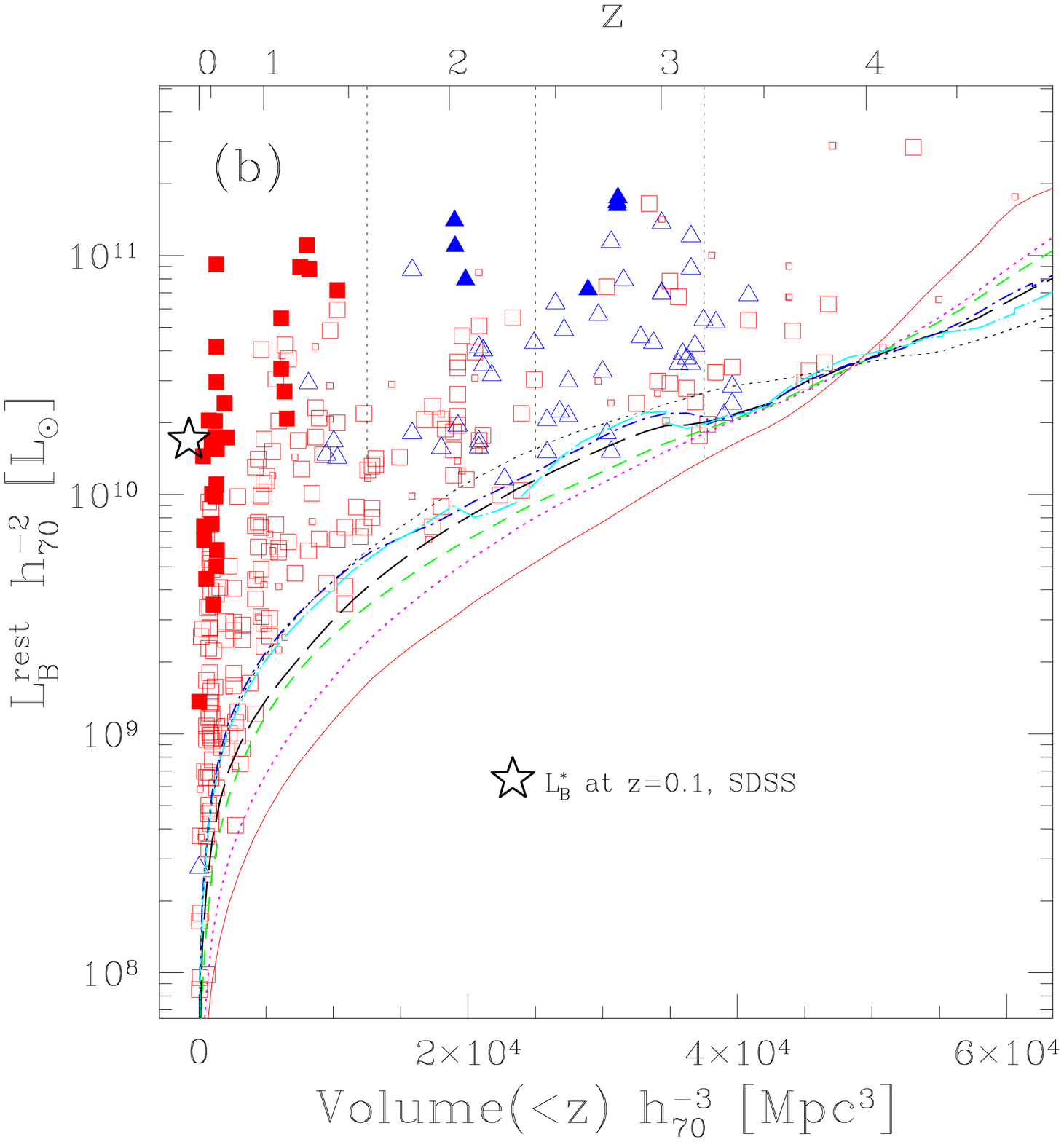}
\end{figure}

\begin{figure}
\epsscale{.80}
\plotone{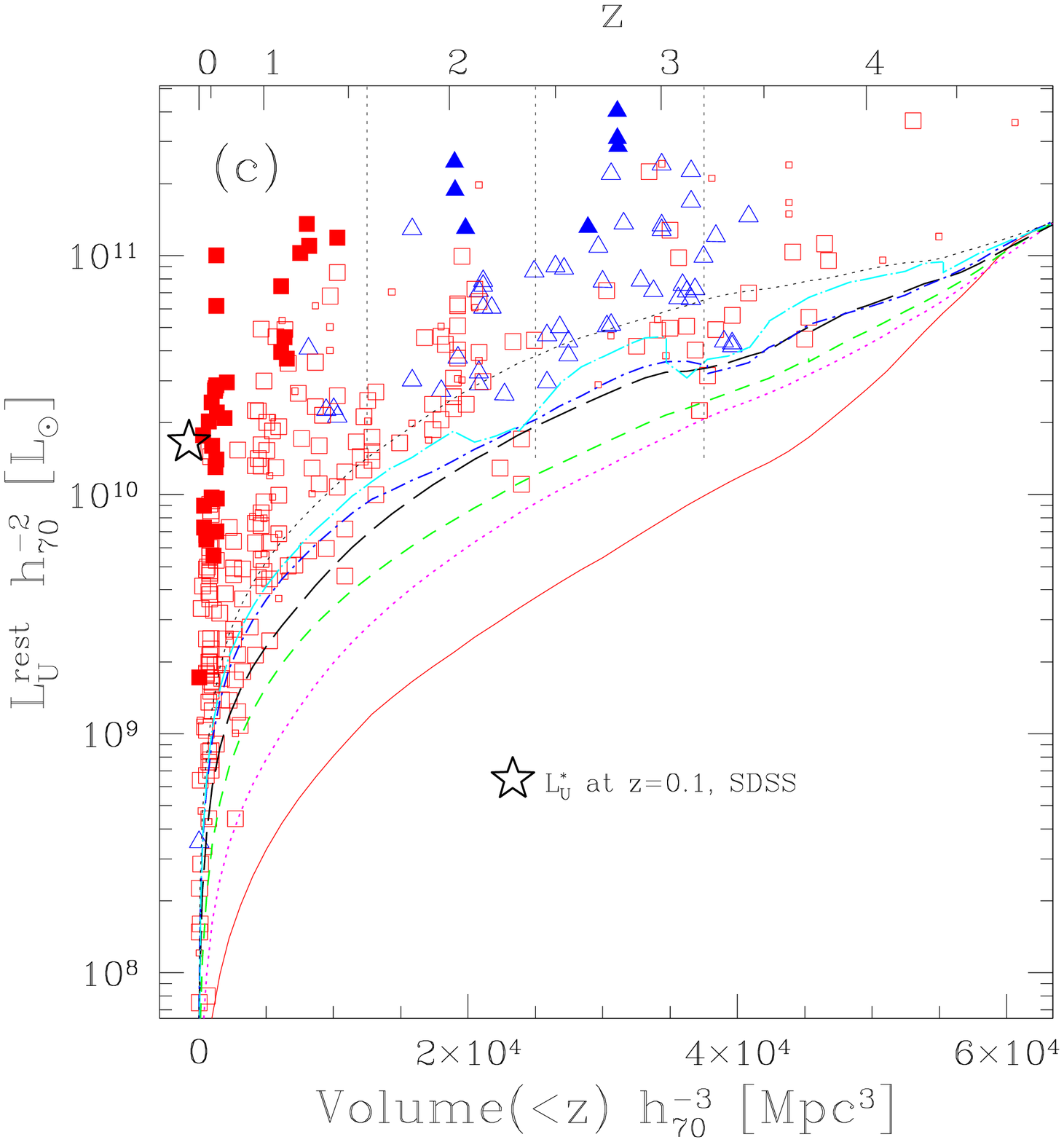}
\end{figure}

\section{The Properties of the Massive Galaxy Population}
\label{results}

 In this section we discuss the use of the \zp and \lrest estimates to
 derive the integrated properties of the population, namely the
 luminosity density, the mean cosmic rest-frame color, the stellar
 mass-to-light ratio \mlstar, and the stellar mass density \rhostar.
 As will be described below, addressing the integrated properties of
 the population reduces many of the uncertainties associated with
 modeling individual galaxies and, in the case of the cosmic color, is
 less sensitive to field-to-field variations.

\subsection{The Luminosity Density}
\label{lumdens}

 Using our \lrest estimates from the \Ksabtot$<25$ galaxies (see
 \S\ref{restlum}), we traced the redshift evolution of the rest-frame
 optically most luminous, and therefore presumably most massive,
 galaxies by measuring the rest-frame luminosity density \jrest of the
 visible stars associated with them.  The results are presented in
 Table~\ref{ldens_tab} and plotted against redshift and elapsed cosmic
 time in Figure~\ref{lumdensfig}.  As our best alternative to a
 selection by galaxy mass, we selected our galaxies in our reddest
 rest-frame band available at $z\sim3$, i.e. the \mv-band.  In
 choosing the $z$ and \lrest regime over which we measured \jrest we
 wanted to push to as high of a redshift as possible with the double
 constraint that the redshifted rest-frame filter still overlapped
 with the \Ks filter and that we were equally complete at all
 considered redshifts.  By choosing an \l{V} threshold,
 \lthreshlam{V}$=1.4 \times 10^{10}$\lumsol, and a maximum redshift of
 $z=3.2$, we could select galaxies down to 0.6~\lstarlocallam{V} with
 constant efficiency regardless of SED type.  We then divided the
 range out to $z=3.2$ into three bins of equal co-moving volume which
 correspond to the redshift intervals 0--1.6, 1.6--2.41, and
 2.41--3.2.

 In a given redshift interval, we estimated \jrest directly from the
 data in two steps.  We first added up all the luminosities of
 galaxies which satisfied our \lthreshlam{V} criteria defined above
 and which had \dzp$/(1 + z_{phot})\leq 0.16$, roughly twice the mean
 disagreement between \zp and \zs (see \S\ref{zpmethod}).  Galaxies
 rejected by our \dzp cut but with \l{V}$>$\lthreshlam{V}, however,
 contribute to the total luminosity although they are not included in
 this first estimate.  Under the assumption that these galaxies are
 drawn from the same luminosity function as those which passed the
 \dzp cut, we computed the total luminosity, including the light from
 the $n_{acc}$ accepted galaxies and the light lost from the $n_{rej}$
 rejected galaxies as
\begin{equation}
{\rm L_{tot}} = {\rm L_{meas}}\times (1 + \frac{n_{rej}}{n_{acc}}).
\end{equation}
 As a test of the underlying assumption of this correction we
 performed a K-S test on the distributions of \Ks magnitudes for the
 rejected and accepted galaxies in each of our three volume bins.  In
 all three redshift bins, the rejected galaxies have \Ksabtot
 distributions which are consistent at the $>90\%$ level with being
 drawn from the same magnitude distribution as the accepted galaxies.
 The total correction per volume bin ranges from $5-10\%$ in every
 bin.  Our results do not change if we omit those galaxies whose
 photometric redshift probability distribution indicates a secondary
 minima in chisquared.

 Uncertainties in the luminosity density were computed by
 bootstrapping from the \Ksabtot$<25$ subsample.  This method does not
 take cosmic variance into account and the errors may therefore
 underestimate the true error, which includes field-to-field variance.

 Redshift errors might effect the luminosity density in a systematic
 way, as they produce a large error in the measured luminosity. This,
 combined with a steep luminosity function can bias the observed
 luminosities upwards, especially at the bright end. This effect can
 be corrected for in a full determination of the luminosity function
 (e.g., Chen et al. 2003), but for our application we estimated the
 strength of this effect by Monte-Carlo simulations. When we used the
 formal redshift errors we obtained a very small bias (6\%), if we
 increase the photometric redshift errors in the simulation to be
 minimally as large as $0.08*(1+z)$, we still obtain a bias in the
 luminosity density on the order of 10\% or less.

 Because we exclude galaxies with faint rest-frame luminosities or low
 apparent magnitudes, and do not correct for this incompleteness, our
 estimates should be regarded as lower limits on the total luminosity
 density.  One possibility for estimating the total luminosity density
 would be to fit a luminosity function as a function of redshift and
 then integrate it over the whole luminosity range.  We don't go faint
 enough at high redshift, however, to tightly constrain the faint-end
 slope $\alpha$.  Because extrapolation of \jrest to arbitrarily low
 luminosities is very dependent on the value of $\alpha$, we choose to
 use this simple and direct method instead.  Including all galaxies
 with \Ksabtot$<25$ raises the \jrest values in the $z=0-1.6$ redshift
 bin by 86\%, 74\%, and 66\% in the \u, \mb, and \mv-bands
 respectively.  Likewise, the \jrest values would increase by 38\%,
 35\%, and 44\% for the $z=1.6-2.41$ bin and would increase by 5\%,
 5\%, and 2\% for the $z=2.41-3.2$ bin, again in the \u, \mb, and
 \mv-bands respectively.

 The dip in the luminosity density in the second lowest redshift bin
 of the (a) and (b) panels of Figure~\ref{lumdensfig} can be traced to
 the lack of intrinsically luminous galaxies at $z \sim 1.5 - 2$
 (\S\ref{restlum}; R01).  The dip is not noticeable in the \u-band
 because the galaxies at $z\sim 2$ are brighter with respect to the
 $z<1.6$ galaxies in the \u-band than in the \mv or \mb-band,
 i.e. they have bluer \ubr colors and \uvr colors than galaxies at
 $z<1.6$.  This lack of rest-frame optically bright galaxies at $z
 \sim 1.5 - 2$ may result from systematics in the \zp estimates, which
 are poorly tested in this regime and where the Lyman break has not
 yet entered the \magu filter, or may reflect a true deficit in the
 redshift distribution of \Ks-band luminous galaxies (see
 \S\ref{restlum}).

 At $z\lesssim 1$ our survey is limited by its small volume.  For this
 reason, we supplement our data with other estimates of \jrest at
 $z\lesssim1$.

 We compared our results with those of the SDSS as follows. First, we
 selected SDSS Main sample galaxies (Strauss et al. 2002) with
 redshifts in the SDSS Early Data Release (Stoughton et al. 2002) in
 the EDR sample provided by and described by \citet{blanton03a}. Using
 the product {\tt kcorrect v1\_16} \citep{blanton03b}, for each galaxy
 we fit an optical SED to the
 $^{0.1}{\!u}^{0.1}{\!g}^{0.1}{\!r}^{0.1}{\!i}^{0.1}{\!z}$ magnitudes,
 after correcting the magnitudes to $z=0.1$ for evolution using the
 results of \citet{blanton03c}.  We projected this SED onto the $UBV$
 filters as described by Bessell (1990) to obtain absolute magnitudes
 in the $UBV$ Vega-relative system. Using the method described in
 \citet{blanton03a} we calculated the maximum volume
 $V_{\mathrm{max}}$ within the EDR over which each galaxy could have
 been observed, accounting for the survey completeness map and the
 flux limit as a function of position. $1/V_{\mathrm{max}}$ then
 represents the number density contribution of each galaxy. From these
 results we constructed the number density distribution of galaxies as
 a function of color and absolute magnitude and the contribution to
 the uncertainties in those densities from Poisson statistics.  While
 the Poisson errors in the SDSS are negligible, cosmic variance does
 contribute to the uncertainties.  For a more realistic error
 estimate, we use the fractional errors on the luminosity density from
 \citet{blanton03c}.  For the SDSS luminosity function, our
 \lthreshlam{V} encompasses 54\% of the total light.

 In Figure~\ref{lumdensfig} we also show the \jrest measurements from
 the COMBO-17 survey \citep{Wolf03}.  We used a catalog with updated
 redshifts and 29471 galaxies at $z<0.9$, of which 7441 had
 \l{V}$>$\lthreshlam{V} (the J2003 catalog; Wolf, C. private
 communication).  Using this catalog we calculated \jrest in an
 identical way to how it was calculated for the FIRES data.  We
 divided the data into redshift bins of $\Delta z=0.2$ and counted the
 light from all galaxies contained within each bin which had
 \l{V}$>$\lthreshlam{V}.  The formal 68\% confidence limits were
 calculated via bootstrapping.  In addition, in
 Figure~\ref{lumdensfig} we indicate the rms field-to-field variations
 between the three spatially distinct COMBO-17 fields.  As also
 pointed out in \citet{Wolf03}, the field-to-field variations dominate
 the error in the COMBO-17 \jrest determinations.

 \citet{Bell03b} point out that uncertainties in the absolute
 calibration and relative calibration of the SDSS and Johnson
 zeropoints can lead to $\lesssim 10\%$ errors in the derived
 rest-frame magnitudes and colors of galaxies.  To account for this,
 we add a 10\% error in quadrature with the formal errors for both the
 COMBO-17 and SDSS luminosity densities.  These are the errors
 presented in Table~\ref{ldens_tab} and Figure~\ref{lumdensfig}.

 In Figure~\ref{lumdensfig}a we also plot the \jrestlam{V} value of
 luminous LBGs determined by integrating the luminosity function of
 \citet{Shap01} to \lthreshlam{V}.  A direct comparison between our
 sample and theirs is not entirely straightforward because the LBGs
 represent a specific class of non-obscured, star forming galaxies at
 high redshift, selected by their rest-frame far UV light.
 Nonetheless, our \jrest determination at $z=2.8$ is slightly higher
 than their determination at $z=3$, indicating either that the HDF-S
 is overdense with respect to the area surveyed by Shapley et al. or
 that we may have galaxies in our sample which are not present in the
 ground-based LBG sample.

 D03 have also measured the luminosity density in the rest-frame
 \mb-band but, because they do not give their luminosity function
 parameters except for their lowest redshift bin, it is not possible
 to overplot their luminosity density integrated down to our \l{V}
 limit.

\ifemulate
   \begin{deluxetable*}{lllllll}
\else
   \begin{deluxetable}{lllllll}
\fi
\tablewidth{0pt}
\tablecaption{Rest-Frame Optical Luminosity Density and Integrated Color}
\tablehead{\colhead{$z$} & \colhead{log \jrestlam{U}} & \colhead{log \jrestlam{B}} &
 \colhead{log \jrestlam{V}} & \colhead{$(U-B)_{rest}$} & \colhead{$(B-V)_{rest}$}\\
 & [$h_{70}~\mathrm{L}_{\odot,U} $Mpc$^{-3}$] &[$h_{70}~\mathrm{L}_{\odot,B} $Mpc$^{-3}$] &[$h_{70}~\mathrm{L}_{\odot,V} $Mpc$^{-3}$]}
\startdata
$0.10\pm0.10$\tablenotemark{a}  &   $7.89^{+0.04}_{-0.05}$ & $7.87^{+0.04}_{-0.05}$ & $7.91^{+0.04}_{-0.05}$ & $0.14^{+0.02}_{-0.02}$ & $0.75^{+0.02}_{-0.02}$ \\
$0.30\pm0.10$\tablenotemark{b} & $7.84^{+0.05}_{-0.05}$ & $7.85^{+0.05}_{-0.05}$ & $7.93^{+0.05}_{-0.05}$ & $0.21^{+0.02 }_{-0.02}$ & $0.84^{+0.01}_{-0.01}$ \\
$0.50\pm0.10$\tablenotemark{b} & $8.01^{+0.04}_{-0.05}$ & $7.99^{+0.04}_{-0.05}$ & $8.01^{+0.04}_{-0.05}$ & $0.16^{+0.01 }_{-0.01}$ & $0.69^{+0.005}_{-0.01}$ \\
$0.70\pm0.10$\tablenotemark{b} & $8.18^{+0.04}_{-0.05}$ & $8.13^{+0.04}_{-0.05}$ & $8.12^{+0.04}_{-0.05}$ & $0.06^{+0.01 }_{-0.01}$ & $0.64^{+0.01}_{-0.01}$ \\
$0.90\pm0.10$\tablenotemark{b} & $8.22^{+0.04}_{-0.05}$ & $8.13^{+0.04}_{-0.05}$ & $8.09^{+0.04}_{-0.05}$ & $-0.04^{+0.004}_{-0.01}$ & $0.55^{+0.005}_{-0.01}$ \\
$1.12^{+0.48}_{-1.12}$\tablenotemark{c} & $8.11^{+0.08}_{-0.09}$ & $8.02^{+0.08}_{-0.08}$ & $8.00^{+0.08}_{-0.09}$ & $-0.04^{+0.03}_{-0.03}$ & $0.61^{+0.02}_{-0.02}$ \\
$2.01^{+0.40}_{-0.41}$\tablenotemark{c} & $8.21^{+0.08}_{-0.10}$ & $8.00^{+0.08}_{-0.10}$ & $7.89^{+0.08}_{-0.09}$ & $-0.34^{+0.04}_{-0.03}$ & $0.38^{+0.04}_{-0.04}$ \\
$2.80^{+0.40}_{-0.39}$\tablenotemark{c} & $8.58^{+0.07}_{-0.08}$ & $8.32^{+0.07}_{-0.08}$ & $8.18^{+0.07}_{-0.08}$ & $-0.44^{+0.04}_{-0.03}$ & $0.29^{+0.04}_{-0.03}$ \\
\enddata
\label{ldens_tab}
\tablecomments{\jrest and rest-frame colors calculated for galaxies with \l{V}$>1.4\times 10^{10}~h_{70}^{-2}~\mathrm{L}_{\odot,V}$.}
\tablenotetext{a}{SDSS}
\tablenotetext{b}{COMBO-17}
\tablenotetext{c}{FIRES}
\ifemulate
   \end{deluxetable*}
\else
   \end{deluxetable}
\fi

\subsubsection{The Evolution of \jrest}

\begin{figure}
\epsscale{1}
\plotone{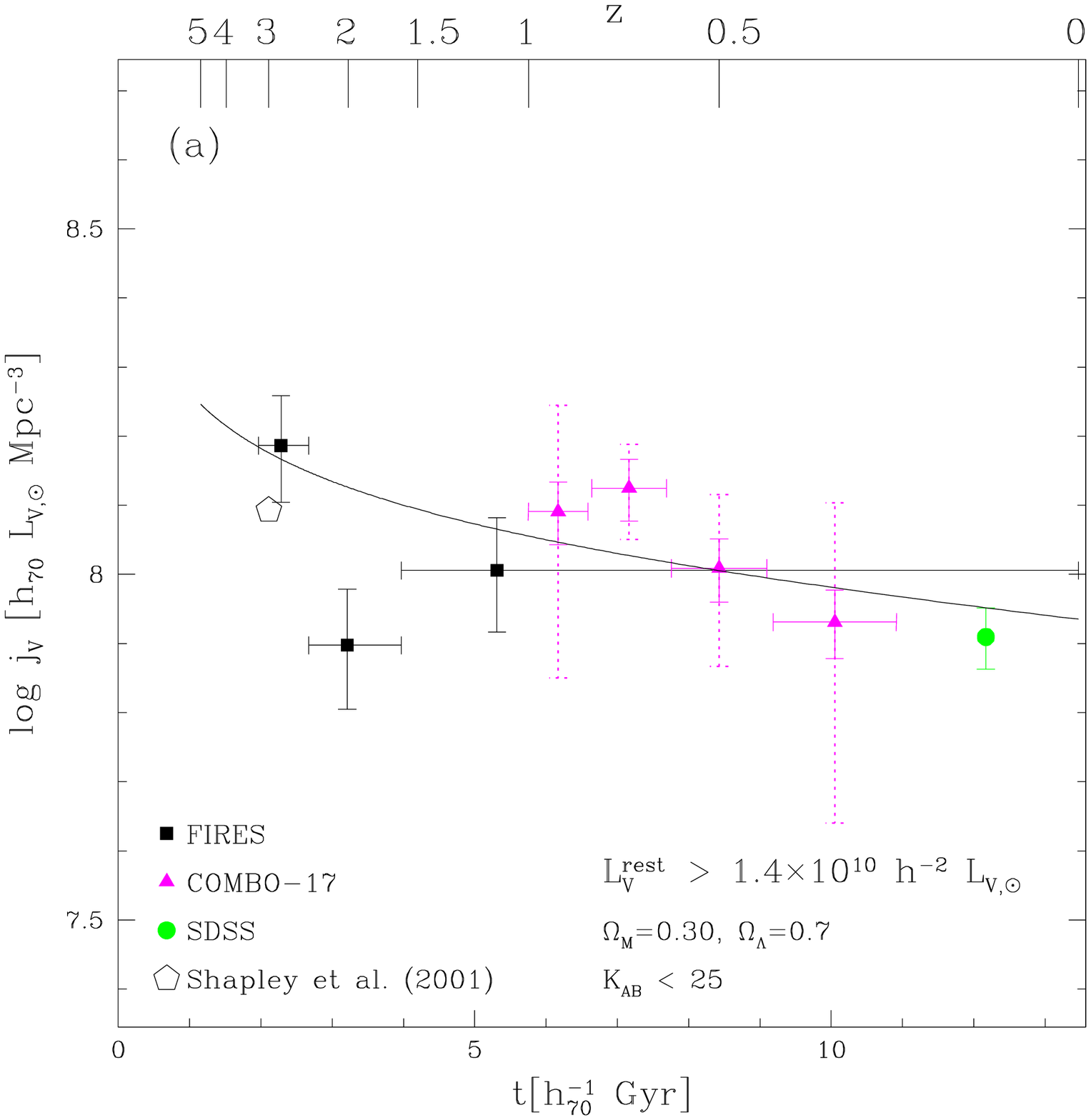}
\caption[Rest-Frame Optical Luminosity Density]{
   The rest-frame optical luminosity density vs. cosmic age and
   redshift from galaxies with \Ksabtot~$<25$ and
   \l{V}$>$\lthreshlam{V}.  For comparison we plot \jrest
   determinations from other surveys down to our \lrest limits.  The
   squares are those from our data, the triangles are from the
   Combo-17 survey \citep{Wolf03}, the circle is that at $z=0.1$ from
   the SDSS (B03), and the pentagon is that from \citet{Shap01}.  The
   dotted errorbars on the COMBO-17 data indicate the rms
   field-to-field variation derived from the three spatially distinct
   COMBO-17 fields.  The solid line is a power law fit to the FIRES,
   COMBO-17, and SDSS data of the form $j^{\rm rest}_\lambda(z) =
   j^{\rm rest}_\lambda(0) * (1 + z)^\beta$.}
\label{lumdensfig}
\end{figure}

 We find progressively stronger luminosity evolution from the \mv to
 the \u-band: whereas the evolution is quite weak in \mv, it is very
 strong in \u.  The \jrest in our highest redshift bin is a factor of
 $1.9\pm0.4$, $2.9\pm0.6$, and $4.9\pm1.0$ higher than the $z=0.1$
 value in \mv, \mb, and \u respectively.  To address the effect of
 cosmic variance on the measured evolution in \jrest we rely on the
 clustering analysis developed for our sample in \citet{daddi03}.
 Using the correlation length estimated at $2<z<4$,
 $r_o=5.5~h_{100}^{-1}$~Mpc, we calculated the expected 1-sigma
 fluctuations in the number density of objects in our two highest
 redshift bins.  Because for our high-$z$ samples the poissonian
 errors are almost identical to the bootstrap errors, we can use the
 errors in the number density as a good proxy to the errors in \jrest.
 The inclusion of the effects of clustering would increase the
 bootstrap errors on the luminosity density by a factor of, at most,
 1.75 downwards and 2.8 upwards.  This implies that the inferred
 evolution is still robust even in the face of the measured
 clustering.  The COMBO-17 data appear to have a slightly steeper rise
 towards higher redshift than our data, however there are two effects
 to remember at this point.  First, our lowest redshift point averages
 over all redshifts $z<1.6$, in which case we are in reasonably good
 agreement with what one would predict from the average of the SDSS
 and COMBO-17 data.  Second, our data may simply have an offset in
 density with respect to the local measurements.  Such an offset
 affects the values of \jrest, but as we will show in \S\ref{color},
 it does not strongly affect the global color estimates.  Nonetheless,
 given the general increase with \jrest towards higher redshifts, we
 fit the changing \jrest with a power law of the form $j^{\rm
 rest}_\lambda(z) = j^{\rm rest}_\lambda(0) * (1 + z)^\beta$.  These
 curves are overplotted in Figure~\ref{lumdensfig} and the best fit
 parameters in sets of ($j^{\rm rest}_\lambda(0)$, $\beta$) are
 ($5.96\times 10^7$,~1.41), ($6.84\times 10^7$,~0.93), and
 ($8.42\times 10^7$,~0.52) in the \u, \mb, and \mv bands respectively,
 where $j^{\rm rest}_\lambda(0)$ has units of
 $\mathrm{h_{70}~L}_{\odot} $Mpc$^{-3}$.  At the same time, it is
 important to remember that our power law fit is likely an
 oversimplification of the true evolution in \jrest.

 The increase in \jrest with decreasing cosmic time can be modeled as
 a simple brightening of \lstar.  Performing a test similar to that
 performed in R01, we determine the increase in \lstarlam{V} with
 respect to \lstarlocallam{V} needed to match the observed increase in
 \jrest from $z=0.1$ to $2.41<z<3.2$, assuming the SDSS Schechter
 function parameters.  To convert between the Schechter function
 parameters in the SDSS bands and those in the \citet{Bess90} filters
 we transformed the \lstarsdss{^{0.1}r} values to the Bessell \mv
 filter using the $(V - ^{0.1}r)$ color, where the color was derived
 from the total luminosity densities in the indicated bands (as given
 in B03). We then applied the appropriate AB to Vega correction
 tabulated in \citet{Bess90}.  Because the difference in
 $\lambda_{eff}$ is small between the two filters in each of these
 colors, the shifts between the systems are less than 5\%.  The
 luminosity density in the \mv-band at $2.41<z<3.2$ is \jrestlam{V}$=
 1.53 \pm 0.26 \times 10^8 {\rm ~h_{70}~L_{\odot}~Mpc^{-3}}$.  Using
 the \mv-band Schechter function parameters for our cosmology,
 $\phi_*^{\rm SDSS}=5.11 \times 10^{-3}{\rm~h_{70}^3~Mpc^{-3}}$,
 $\alpha^{\rm SDSS}=-1.05$, and \lstarsdss{V}$=2.53\times 10^{10}{\rm
 ~h_{70}^{-2}~L_{\odot}}$, we can match the increase in \jrestlam{V}
 if \lstarlam{V} brightens by a factor of 1.7 out to $2.41<z<3.2$.

\begin{figure}
\plotone{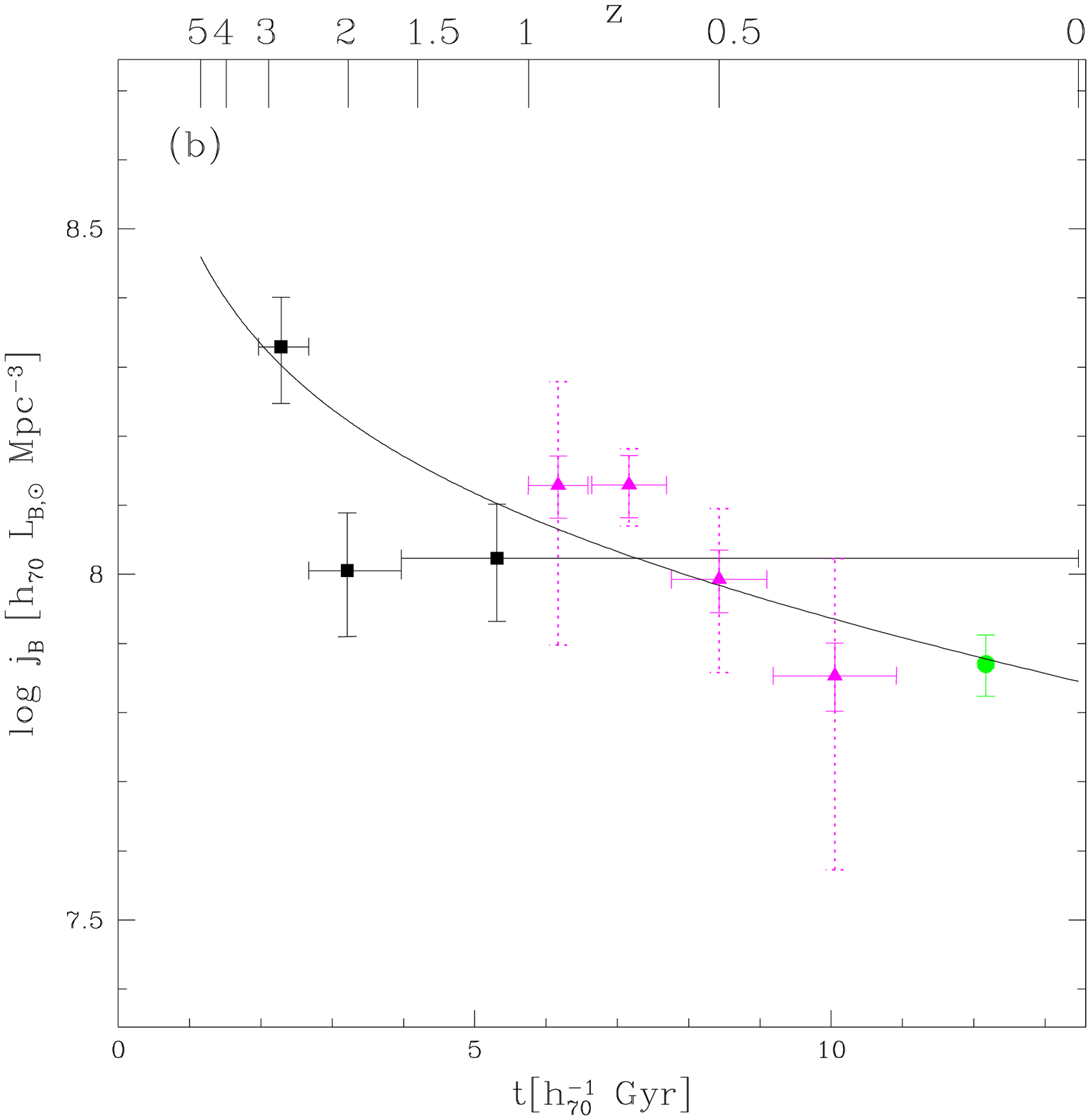}
\end{figure}

\begin{figure}
\plotone{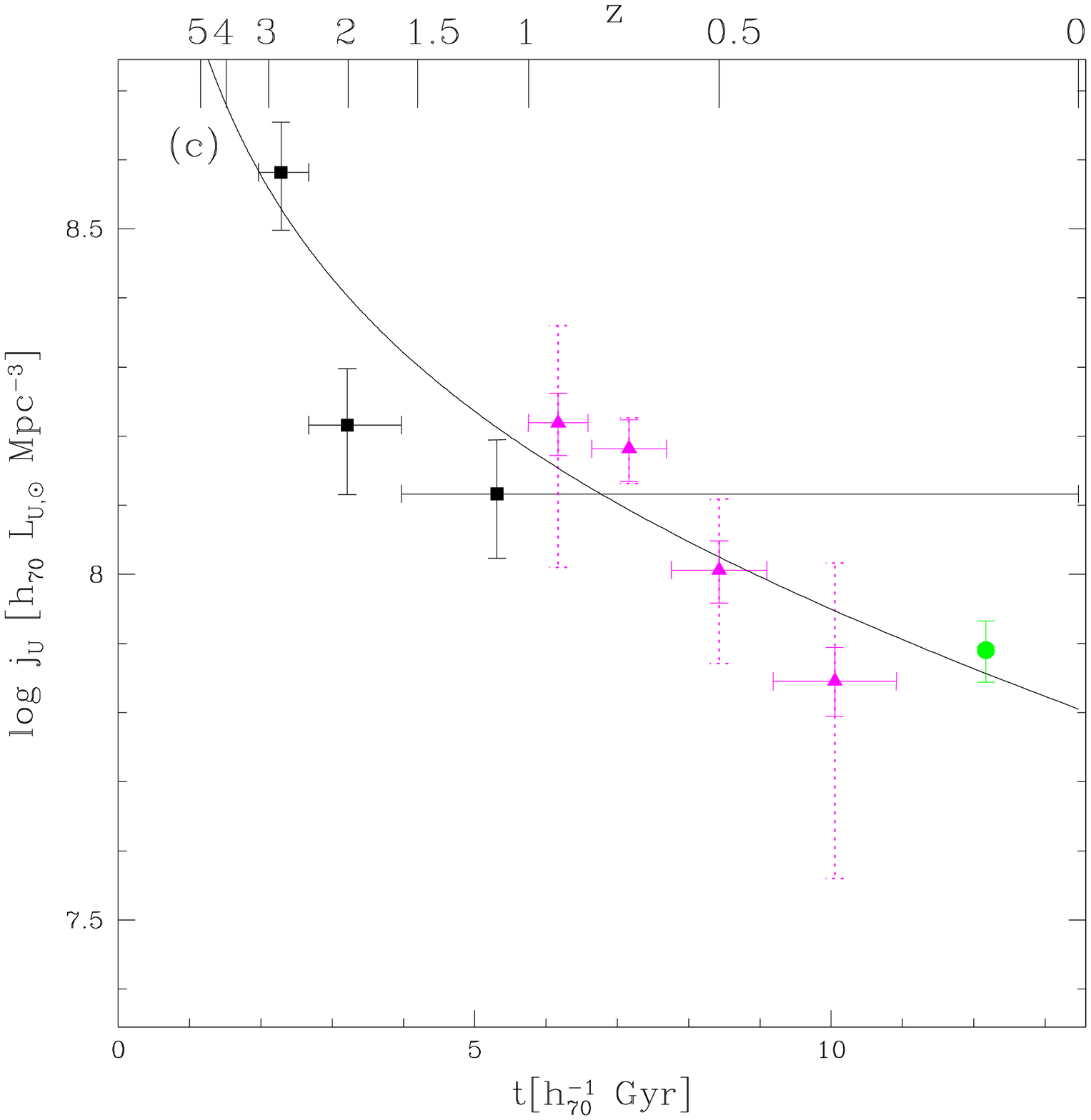}
\end{figure}

\subsection{The Cosmic Color}
\label{color}

 Using our measures of \jrest we estimated the cosmic rest-frame color
 of all the visible stars which lie in galaxies with \l{V}$> 1.4
 \times 10^{10}$\lumsol.  We derived the mean cosmic \ubr and \bvr by
 using the \jrest estimates from the previous section with the
 appropriate magnitude zeropoints.  The measured colors for the FIRES
 data, the COMBO-17 data, and the SDSS data are given in
 Table~\ref{ldens_tab}.  Emission line corrected \ubr colors may be
 calculated by applying Equation~\ref{emiseq} to the values in the
 table.  For the FIRES and COMBO-17 data, uncertainty estimates are
 derived from the same bootstrapping simulation used in
 \S~\ref{lumdens}.  In this case, however, the COMBO-17 and SDSS
 errorbars do not include an extra component from errors in the
 transformation to rest-frame luminosities, since these transformation
 errors may be correlated in a non-trivial way.

\begin{figure}
\epsscale{0.9}
\plotone{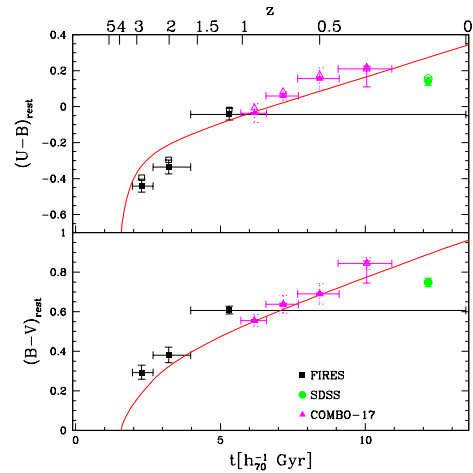}
\caption[Color-redshift relation of the universe]{
 The evolution of the cosmic color plotted against redshift and cosmic
 time for our data in addition to data from other $z\lesssim 1$
 surveys.  The squares are those from our data, the triangles are from
 the Combo-17 survey \citep{Wolf03}, and the circle is that at $z=0.1$
 from the SDSS (B03).  The open symbols indicate the empirical
 emission line correction to the integrated colors derived using the
 spectroscopic and photometric data from the NFGS \citep{Jansen00b}.
 The dotted errorbars on the COMBO-17 data indicate the field-to-field
 variation.  Note that the integrated rest-frame color is much more
 stable than \jrest against field-to-field variations. The COMBO-17
 data point at $z=0.3$ has been given a color errorbar of 0.1 in the
 blueward direction and an open symbol to reflect the possible
 systematic biases resulting from their very small central apertures.
 We also overplot a model with an exponentially declining SFH with
 $\tau=6Gyr$, $E(B-V)=0.35$, and $z_{start}=4.0$ assuming a
 \citet{Calz00} extinction law.}
\label{colzfig}
\end{figure}

 The bluing with increasing redshift which could have been inferred
 from Figure~\ref{lumdensfig} is seen explicitly in
 Figure~\ref{colzfig}.  The color change towards higher redshift
 occurs more smoothly than the evolution in \jrest, with our FIRES
 data meshing nicely with the COMBO-17 data.  It is immediately
 apparent that the rms field-to-field errors for the COMBO-17 data are
 much less than the observed trend in color, in contrast to
 Figure~\ref{lumdensfig}.  This explicitly shows that the integrated
 color is much less sensitive than \jrest to field-to-field
 variations, even when such variations may dominate the error in the
 luminosity density.  The COMBO-17 data at $z\lesssim 0.4$ are also
 redder than the local SDSS data, possibly owing to the very small
 central apertures used to measure colors in the COMBO-17 survey.  The
 colors in the COMBO-17 data were measured with the package
 \texttt{MPIAPHOT} using using the peak surface brightness in images
 smoothed all to identical seeing ($1\farcs5$).  Such small apertures
 were chosen to measure very precise colors, not to obtain global
 color estimates.  Because of color gradients, these small apertures
 can overestimate the global colors in nearby well resolved galaxies,
 while providing more accurate global color estimates for the more
 distant objects.  Following the estimates of this bias provided by
 \citet{Bell03b}, we increased the color errorbars on the blue side to
 0.1 for the $z<0.4$ COMBO-17 data.  It is encouraging to see that the
 color evolution is roughly consistent with a rather simple model and
 that it is much smoother than the luminosity density evolution, which
 is more strongly affected by cosmic variance.
 
 We interpreted the color evolution as being primarily driven by a
 decrease in the stellar age with increasing redshift.  Applying the
 \bvr dependent emission line corrections inferred from local samples
 (See \S\ref{emis}), we see that the effect of the emission lines on
 the color is much less than the magnitude of the observed trend.  We
 can also interpret this change in color as a change in mean cosmic
 \mlstar with redshift.  In this picture, which is true for a variety
 of monotonic SFHs and extinctions, the points at high redshift have
 lower \mlstar than those at low redshift.  At the same time, however,
 the evolution in \jrestlam{V} with redshift is quite weak.  Taken
 together this would imply that the stellar mass density \rhostar is
 also decreasing with increasing redshift.  We will quantify this in
 \S\ref{mass}.

\begin{figure}
\epsscale{.85}
\plotone{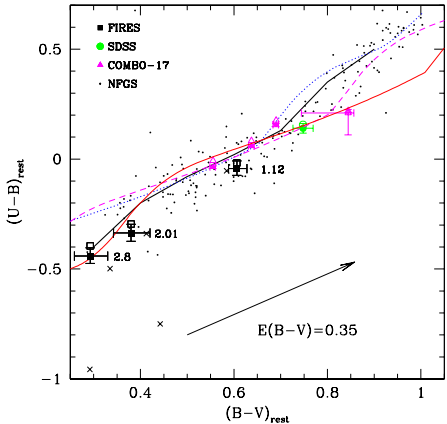}
\caption[Color-Color relation of the universe]{
 The \ubr vs. \bvr at $z=$1.12, 2.01, and 2.8 of all the relatively
 unobscured stars in galaxies with \l{V}$>1.4 \times 10^{10}$\lumsol.
 The thick solid black line is the local relation derived by
 \citet{Lar78} from nearby morphologically normal galaxies.  The large
 symbols are identical to those in Figure~\ref{colzfig}.  For clarity
 we do not plot the field-to-field errorbars for the COMBO-17 data.
 The small solid points are the colors of nearby galaxies from the
 NFGS \citep{Jansen00a}, which have been corrected for emission lines.
 The small crosses are the NFGS galaxies which harbor AGN.  The thin
 tracks correspond to an exponentially declining SFH with a timescale
 of 6~Gyr.  The tracks were created using a \citet{Sal55} IMF and the
 2000 version of the \citet{BC93} models.  The dotted track has no
 extinction, the dashed track has been reddened by $E(B-V)=0.15$, and
 the thin solid track has been reddened by $E(B-V)=0.35$, using the
 \citet{Calz00} extinction law.  The black arrow indicates the
 reddening vector applied to the solid model track.  The emission line
 corrected data lie very close the track defined by observations of
 local galaxies and the agreement with the models demonstrates that
 simple SFHs can be used to reproduce the integrated colors from
 massive galaxies at all redshifts.  }
\label{colcolfig}
\end{figure}

 To show how our mean cosmic \ubr and \bvr colors compare to those of
 morphologically normal nearby galaxies, we overplot them in
 Figure~\ref{colcolfig} on the locus of nearby galaxies from
 \citet{Lar78}.  The integrated colors, at all redshifts, lie very
 close to the local track, which \citet{Lar78} demonstrated is easily
 reproduceable with simple monotonically declining SFHs and which is
 preserved in the presence of modest amounts of reddening, which moves
 galaxies roughly parallel to the locus.  In fact, correcting our data
 for emission lines moved them even closer to the local track.  While
 we have suggested that \mlstar decreases with decreasing color, if we
 wish to actually quantify the \mlstar evolution from our data we must
 first attempt to find a set of models which can match our observed
 colors and which we will later use to convert between the color and
 \mlstarlam{V}.  We overplot in Figure~\ref{colcolfig} two model
 tracks corresponding to an exponential SFH with $\tau=6$ Gyr and with
 $E(B-V)=0$, 0.15, and 0.35 (assuming a \citet{Calz00} reddening law).
 These tracks were calculated using the 2000 version of the
 \citet{BC93} models and have $Z=Z_\odot$ and a \citet{Sal55} IMF with
 a mass range of 0.1-120$M_\odot$.  Other exponentially declining
 models and even a constant star forming track all yield similar
 colors to the $\tau=6$ Gyr track.  The measured cosmic colors at
 $z<1.6$ are fairly well approximated either of the reddening models.
 At $z>1.6$, however, only the $E(B-V)=0.35$ track can reproduce the
 data.  This high extinction is in contrast to the results of P01 and
 \citet{Shap01} who found a mean reddening for LBGs of $E(B-V)\sim
 0.15$.  SY98 and \citet{Thom01}, however, measured extinctions on
 this order for galaxies in the HDF-N, although the mean extinction
 from \citet{Thom01} was lower at $z>2$.  The amount of reddening in
 our sample is one of the largest uncertainty in deriving the \mlstar
 values, nonetheless, our choice of a high extinction is the only
 allowable possibility given the integrated colors of our high
 redshift data.

 Although this figure demonstrates that the measured colors can be
 matched, at some age, by this simple $E(B-V)=0.35$ model, we must
 nevertheless investigate whether the evolution of our model colors
 are also compatible with the evolution in the measured colors.  This
 is shown by the track in Figure~\ref{colzfig}.  We have tried
 different combinations of $\tau$, $E(B-V)$, and $z_{start}$, but have
 not been able to find a model which fits the data well at all
 redshifts.  The parameterized SFR($z$) curve of \citet{Cole01} also
 provided a poor fit to the data.  Given the large range of possible
 parameters, our data may not be sufficient to well constrain the SFH.

\subsection{Estimating \mlstarlam{V} and The Stellar Mass Density}
\label{mass}

 In this subsection we describe the use of our mean cosmic color
 estimates to derive the mean cosmic \mlstarlam{V} and the evolution
 in \rhostar.

 The main strength of considering the luminosity density and
 integrated colors of the galaxy population, as opposed to those of
 individual galaxies, lies in the simple and robust ways in which
 these global values can be modeled.  When attempting to derive the
 SFHs and stellar masses of individual high-redshift galaxies, the
 state-of-the-art models for the broad-band colors only consider
 stellar populations with at most two separate components (SY98; P01;
 Shapley et al. 2001).  Using their stellar population synthesis
 modeling, \citet{Shap01} proposes a model in which LBGs likely have
 smooth SFHs.  On the other hand, SY98 concluded that they may only be
 seeing the most recent episode of star formation and that LBGs may
 indeed have bursting SFHs.  This same idea was supported by P01 and
 \citet{Ferg02} using much deeper NIR data.  When using similar simple
 SFHs to model the cosmic average of the galaxy population, a more
 self-consistent approach is possible.  While individual galaxies may,
 and probably do, have complex SFHs, the mean SFH of all galaxies is
 much smoother than that of individual ones.

\begin{figure}
\epsscale{1}
\plotone{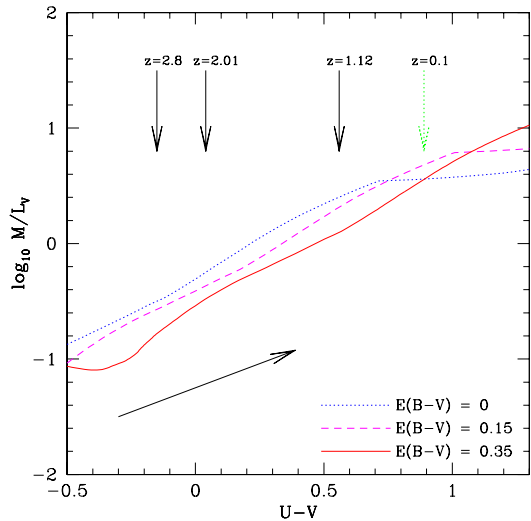}
\caption[\mlstarlam{V} vs. \uv]{
  The relation between \uv and \mlstarlam{V} for a model track with an
  exponential timescale of 6 Gyr.  The dotted line is for a model with
  $E(B-V)=0$, the dashed line for a model with $E(B-V)=0.15$, and the
  solid line is for a model reddened by $E(B-V)=0.35$ (using a
  Calzetti extinction law), which we adopt for our \mlstarlam{V}
  conversions.  The vertical solid arrows indicate the colors of the
  three FIRES data points, the vertical dotted arrow indicates the
  color of the SDSS data, and the diagonal solid arrow indicates the
  vector used to redden the $E(B-V)=0$ model to $E(B-V)=0.35$.  The
  labels above the vertical arrows correspond to the redshifts of the
  FIRES and SDSS data. }
\label{ml_colrelfig}
\end{figure}

 Encouraged by the general agreement between the measured colors and
 the simple models, we attempted to use this model to constrain the
 stellar mass-to-light ratio \mlstarlam{V} in the rest-frame \mv-band,
 by taking advantage of the relation between color and
 log$_{10}$\mlstar found by \citet{Bell01}.  For monotonic SFHs, the
 scatter of this relation remains small in the presence of modest
 variations in the reddening and metallicity because these effects run
 roughly parallel to the mean relation.  Using the $\tau=6$ Gyr
 exponentially declining model, we plot in Figure~\ref{ml_colrelfig}
 the relation between \uvr and \mlstarlam{V} for the $E(B-V)=0$, 0.15,
 and 0.35 models.  It is seen that dust extinction moves objects
 roughly parallel to the model tracks, reddening their colors, but
 making them dimmer as well and hence increasing \mlstarlam{V}.
 Nonetheless, extinction uncertainties are a major contributor to our
 errors in the determination of \mlstarlam{V}.  We chose to derive
 \mlstarlam{V} from the \uvr color instead of from the \bvr color
 because at blue colors, where our high redshift points lie, \bvr
 derived \mlstarlam{V} values are much more sensitive to the exact
 value of the extinction.  This behavior likely stems from the fact
 that the \uvr color spans the Balmer/4000\ang break and hence is more
 age sensitive than \bvr. At the same time, while \ubr colors are even
 less sensitive to extinction than \uvr, they are more susceptible to
 the effects of bursts.

 We constructed our relation using a \citet{Sal55} IMF\footnote{We do
 not attempt to model an evolving IMF although evidence for a
 top-heavy IMF at high redshifts has been presented by
 \citet{Ferg02}}.  The adoption of a different IMF would simply change
 the zeropoint of this curve, leaving the relative \mlstar as a
 function of color, however, unchanged.  As discussed in
 \S\ref{color}, this model does not fit the redshift evolution of the
 cosmic color very well.  Nonetheless, the impact on our \mlstarlam{V}
 estimates should not be very large, since most smooth SFHs occupy
 very similar positions in the \mlstarlam{V} vs. $U-V$ plane.

\begin{figure}
\plotone{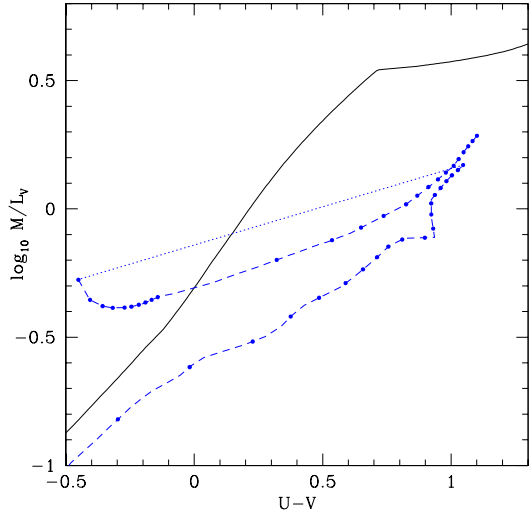}
\caption[\mlstarlam{V} vs. \uv for SFHs with an initial burst]{
 The effect of an early burst of star formation on the relation
 between \uv and \mlstarlam{V}.  The relation between \uv and
 \mlstarlam{V} for a model track with an exponential timescale of 6
 Gyr is show by the solid line.  We also show a track for a SFH which
 includes a 50 Myr burst at $t=0$ followed by a gap of 2 Gyr and then
 a constant SFR rate for 1 Gyr thereafter, where the fraction of mass
 formed in the burst is 0.5.  The track continues for a total time of
 4.5 Gyr.  The dots are placed at 100 Myr intervals and the dotted
 section of the line indicates the very rapid transition in color
 caused by the onset of the second period of star formation.  Both
 tracks have the same extinction. }
\label{ml_colrel_trackcomp}
\end{figure}

 This relation breaks down in the presence of more complex SFHs.  We
 demonstrate this in Figure~\ref{ml_colrel_trackcomp} where we plot
 the $\tau=6$ Gyr track and a second track whose SFH is comprised of a
 50 Myr burst at $t=0$, a gap of 2 Gyr, and a constant SFR rate for
 1~Gyr thereafter, where 50\% of the mass is formed in the burst.  It
 is obvious from this figure that using a smooth model will cause
 errors in the \mlstarlam{V} estimate if the galaxy has a SFR which
 has an early peak in the SFH.  At blue colors, such a early burst of
 SFR will cause an underestimate of \mlstarlam{V}, a result similar to
 that of P01 and D03.  At red colors, however, \mlstarlam{V} would be
 overestimated with the exact systematic offset as a function of color
 depending strongly on the detailed SFH, i.e. the fraction of mass
 formed in the burst, the length of the gap, and the final age of the
 stellar population.  

\begin{figure}
\plotone{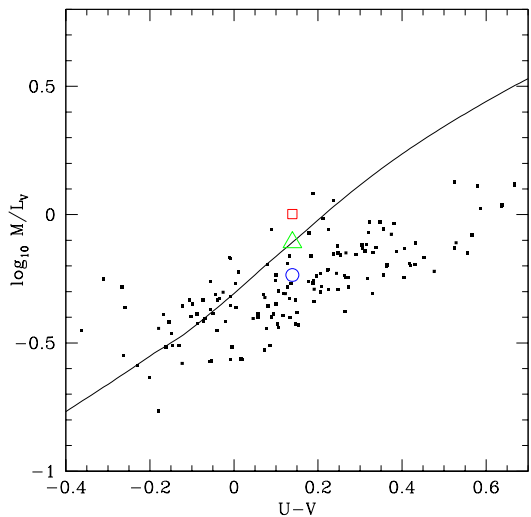}
\caption[\mlstarlam{V} vs. \uv for Bursting SFHs]{
 A comparison of different measures of the global \mlstarlam{V} for a
 mock catalog of galaxies with bursting SFHs.  The solid line represents
 the relation between \uv and \mlstarlam{V} for a model track with an
 exponential timescale of 6 Gyr.  The black dots show the true M/L's
 of the model starbursting galaxies, as described in the text; the open
 circle shows the true luminosity weighted ${\rm M_{tot}}/{\rm
 L_{tot}}$ of the mock galaxies.  The square shows the luminosity
 weighted \mlstarlam{V} derived by applying the simple model to the
 individual galaxies - in this case, the mean ${\rm M_{tot}}/{\rm
 L_{tot}}$ is overestimated by 70\%.  The triangle is the ${\rm
 M_{tot}}/{\rm L_{tot}}$ derived from the luminosity weighted mean
 color (or $(U - V)_{\rm tot}$) of the model galaxies.  It
 overestimates ${\rm M_{tot}}/{\rm L_{tot}}$ by only 35\%. }
\label{ml_colrel_burstfig}
\end{figure}

 The models show that the method may over- or underestimate the true
 stellar mass-to-light ratio if the galaxies have complex SFHs.  It is
 important to quantify the errors on the global \mlstarlam{V} based on
 the mean \uvr color and how these errors compare to those when
 determining the global \mlstarlam{V} value from individual
 \mlstarlam{V} estimates.  To make this comparison we constructed a
 model whose SFH consist of a set of 10 Myr duration bursts separated
 by 90 Myr gaps.  We drew galaxies at random from this model by
 randomly varying the phase and age of the burst sequence, where the
 maximum age was 4 Gyr.  Next we estimated the total mass-to-light
 ratios of this sample by two different methods; first we determined
 the \mlstarlam{V} for the galaxies individually assuming the simple
 relation between color and mass-to-light ratio, and we took the
 luminosity weighted mean of the individual estimates to obtain the
 total \mlstarlam{V}.  This point is indicated by a large square in
 Figure~\ref{ml_colrel_burstfig} and overestimates the total
 \mlstarlam{V} by $\sim 70$\%.  Next we first add the light of all the
 galaxies in both \u and \mv, then use the simple relation between
 color and \mlstarlam{V} to convert the integrated \uvr into a
 mass-to-light ratio.  This method overestimates \mlstarlam{V} by much
 less, $\sim 35$\%.  This comparison shows clearly that it is best to
 estimate the mass using the integrated light. This is not unexpected;
 the star formation history of the universe as a whole is more regular
 than the star formation history of individual galaxies. If enough
 galaxies are averaged, the mean star formation history is naturally
 fairly smooth.

\begin{figure}
\epsscale{0.9}
\plotone{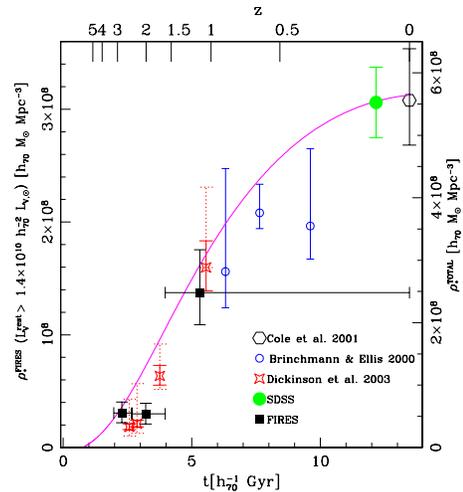}
\caption[Stellar Mass Density]{
 The build-up of the stellar mass density as a function of redshift.
 The solid points are for galaxies with \l{V}$>1.4 \times
 10^{10}$\lumsol and were derived by applying the $E(B-V)=0.35$
 relation in Figure~\ref{ml_colrelfig} to the \uvr colors and \jrest
 measurements from the FIRES (solid squares) and SDSS data (solid
 circle).  The y-axis scale on the left side corresponds to the
 \rhostar values for these points.  The hollow points show the total
 stellar mass density measurements from the one-component models in
 the HDF-N (D03; hollow stars; calculated assuming solar metallicity),
 the CFRS (Brinchmann \& Ellis 2000; hollow circles), and the 2dFGRS +
 2MASS (Cole et al. 2001; hollow hexagon).  The dotted errorbars on
 the D03 points reflect the systematic mass uncertainties resulting
 from metallicity and SFH changes.  The y-axis scale on the right hand
 side corresponds to the \rhostar estimates for these points.  The
 relative scaling of the two axes was adjusted so that our SDSS
 \rhostar estimate was at the same height as the total \rhostar
 estimate of Cole et al.  The solid curve is an integral of the
 SFR($z$) from \citet{Cole01} which has been fit to extinction
 corrected data at $z\lesssim4$.  }
\label{massdensfig}
\end{figure}

 Using the relationship between color and \mlstarlam{V} we convert our
 \uvr and \jrestlam{V} measurements to stellar mass density estimates
 \rhostar.  The resulting \rhostar values are plotted vs. cosmic time
 in Figure~\ref{massdensfig}.  We have included points for the SDSS
 survey created in an analogous way to those from this work,
 i.e. using the \mlstarlam{V} derived from the rest-frame color and
 multiplying it by \jsdss{V} for all galaxies with
 \l{V}$>$\lthreshlam{V}.  The \rhostar estimates are listed in
 Table~\ref{rhostar_tab}.  We have derived the statistical errorbars
 on the \rhostar estimates by creating a Monte-Carlo simulation where
 we allowed \jrestlam{V} and \uvr (and hence \mlstarlam{V}) to vary
 within their errorbars.  We then took the resulting distribution of
 \rhostar values and determined the 68\% confidence limits.  As an
 estimate of our systematic uncertainties corresponding to the method
 we also determined \mlstarlam{V} from the \ubr and \bvr data using an
 identical relation as for the \uvr to \mlstarlam{V} conversion.  The
 \ubr derived \mlstarlam{V} values were different from the \uvr values
 by a factor of 1.02, 0.80, 0.95, and 1.12 for the $z=$0.1, 1.12,
 2.01, and 2.8 redshift bins respectively.  Likewise the \bvr
 determined \mlstarlam{V} values changed by a factor of 0.99, 1.11,
 1.15, and 0.80 with respect to the \uvr values.  While the \uvr
 values are very similar to those derived from the other colors, the
 \uvr color is less susceptible to dust uncertainties than the \bvr
 data and less susceptible to the effects of bursts than the \ubr
 data.  

 The derived mass density rises monotonically by a factor of $\sim 10$
 all the way to $z\sim 0.1$, with our low redshift point meshing
 nicely with the local SDSS point.

\ifemulate
   \begin{deluxetable*}{lcc}
\else
   \begin{deluxetable}{lcc}
\fi
\tablewidth{0pt}
\tablecaption{\mlstarlam{V} and Stellar Mass Density Estimates}
\tablehead{\colhead{$z$} & \colhead{log \mlstarlam{V}} & \colhead{log \rhostar} \\
 & $[\frac{\mathrm{M_{\odot}}}{\mathrm{L_{\odot}}}] $ & [$h_{70}~\mathrm{M_{\odot}} $Mpc$^{-3}$] }
\startdata
$0.1\pm0.1$\tablenotemark{a} & $0.54_{-0.03}^{+0.03}$ & $8.49_{-0.05}^{+0.04}$ \\
$1.12^{+0.48}_{-1.12}$\tablenotemark{b} & $0.13_{-0.06}^{+0.07}$ & $8.14_{-0.10}^{+0.11}$\\
$2.01^{+0.40}_{-0.41}$\tablenotemark{b} & $-0.42_{-0.10}^{+0.09}$ & $7.48_{-0.16}^{+0.12}$\\
$2.80^{+0.40}_{-0.39}$\tablenotemark{b} & $-0.70_{-0.12}^{+0.11}$ & $7.49_{-0.14}^{+0.12}$\\
\enddata
\label{rhostar_tab}
\tablenotetext{a}{SDSS}
\tablenotetext{b}{FIRES}
\ifemulate
   \end{deluxetable*}
\else
   \end{deluxetable}
\fi

\section{Discussion}
\label{discuss}

\subsection{Comparison with other Work}

 Figure~\ref{massdensfig} shows a consistent picture of the build-up
 of stellar mass, both for the luminous galaxies and the total galaxy
 population.  It is remarkable that the results from different authors
 appear to agree well given that the methods to derive the densities
 were different and that the fields are very small.

 We compared our results to the total mass estimates of other authors
 in Figure~\ref{massdensfig}.  In doing this we must remember, because
 of our \lthresh cut, that we are missing significant amounts of light,
 and hence, mass.  Assuming the SDSS luminosity function parameters,
 we lose 46\% of the light at $z=0$.  At $z=2.8$, however, we inferred
 a brightening of \lstarlam{V} by a factor of 1.7, implying that we go
 further down the luminosity function at high redshift, sampling a
 larger fraction of the total starlight.  If we apply this brightening
 to the SDSS \lstarlam{V} we miss 30\% of the light below our
 luminosity threshold at $z=2.8$.  Hence, the fraction of the total
 starlight contained in our sample is rather stable as a function of
 redshift.  To graphically compare our data to other authors we have
 scaled the two different axes in Figure~\ref{massdensfig} so that our
 derivation of the SDSS \rhostar is at the same height of the total
 \rhostar estimate of \citet{Cole01}.  At $z<1$ we compared our mass
 estimates to those of \citet{Brinch00}.  Following D03, we have
 corrected their published points to total masses by correcting them
 upwards by 20\% to account for their mass incompleteness.  The
 fraction of the total stars formed at $z<1$ agrees well between our
 data and that of SDSS and Brinchmann \& Ellis.  At $z>0.5$, we
 compared our results to those of D03.  D03 calculated the total mass
 density, using the integrated luminosity density in the rest-frame
 B-band coupled with M/L measurements of individual galaxies.  The
 fractions of the total stars formed in our sample (60\%, 13\%, and
 9\% at $z=$1.12, 2.01, and 2.8) are almost twice as high as those of
 D03.  The results, however, are consistent within the errors.

 We explored whether field-to-field variations may play a role in the
 discrepancy between the two datasets.  D03 studied the HDF-N, which
 has far fewer "red" galaxies than HDF-S (e.g., Labb\'e et al. 2003,
 Franx et al, 2003).  If we omit the $J-K$ selected galaxies found by
 \citet{Franx03} in the HDF-S, the formal \mlstarlam{V} decreases to
 45\% and 43\% of the total values and the mass density decreases to
 57\% and 56\% of the total values in the $z=2.01$ and $z=2.8$ bins
 respectively, bringing our data into better agreement with that from
 D03.  This reinforces the earlier suggestion by \citet{Franx03} that
 the $J-K$ selected galaxies contribute significantly to the stellar
 mass budget.

 The errors in both determinations are dominated by systematic
 uncertainties, although our method should be less sensitive to bursts
 than that of D03 as it uses the light integrated integrated over the
 galaxy population.

 We note that \citet{fontana03} have also measured the stellar mass
 density in the HDF-S using a catalog derived from data in common with
 our own.  They find a similar, although slightly smaller evolution in
 the stellar mass density, consistent with our result to within the
 uncertainties.

\subsection{Comparison with SFR($z$)}

 We can compare the derived stellar mass to the mass expected from
 determinations of the SFR as a function of redshift. We use the curve
 by \citet{Cole01}, who fitted the observed SFR as determined from
 various sources at $z\lesssim 4$.  To obtain the curve in
 Figure~\ref{massdensfig} we integrated the SFR($z$) curve taking into
 account the time dependent stellar mass loss derived from the 2000
 version of the \citet{BC93} population synthesis models.

 We calculated a reduced $\chi^2$ of 4.3 when comparing all the data
 to the model.  If, however, we omit the \citet{Brinch00} data, the
 reduced $\chi^2$ decreases to 1.8, although the results at $z>2$ lie
 systematically below the curve. This suggests that some systematic
 errors may play a role, or that the curve is not quite correct.  The
 following errors can influence our mass density determinations:

 -Dusty, evolved populations: it is assumed that the dust is mixed in
 a simple way with the stars, leading to a Calzetti extinction curve.
 If the dust is distributed differently, e.g., by having a very
 extincted underlying evolved population, or by having a larger R
 value, the current assumptions lead to a systematic underestimate of
 the mass.  If an underlying, extincted evolved population exists, it
 would naturally explain the fact that the ages of the Lyman-break
 galaxies are much younger than expected (e.g., P01, Ferguson et
 al. 2002).  There may also be galaxies which contribute significantly
 to the mass density but are so heavily extincted that they are
 undetected, even with our very deep \Ks-band data.  If such objects
 are also actively forming stars, they may be detectable with deep
 submillimeter observations or with rest-frame NIR observations from
 SIRTF.

 -Cosmic variance: the two fields which have been studied are very
 small. Although we use a consistent estimate of clustering from
 \citet{daddi03}, red galaxies make up a large fraction of the mass
 density in our highest redshift bins.  Since red galaxies have a very
 high measured clustering from $z\sim1$ (e.g., Daddi et al. 2000,
 Mccarthy et al. 2001) up to possibly $z\sim3$ \citep{daddi03}, large
 uncertainties remain.

 -Evolving Initial Mass Function: the light which we see is mostly
 coming from the most massive stars present, whereas the stellar mass
 is dominated by low mass stars.  Changes in the IMF would immediately
 lead to different mass estimates but if the IMF everywhere is
 identical (as we assume), then the relative masses should be robust.
 If the IMF evolves with redshift, however, systematic errors in the
 mass estimate will occur.

 -A steep galaxy mass function at high redshift: if much of the UV
 light which is used to measure the SFR at high redshifts comes from
 small galaxies which would fall below our rest-frame luminosity
 threshold then we may be missing significant amounts of stellar mass.
 Even the mass estimates of D03, which were obtained by integrating
 the luminosity function, are very sensitive to the faint end
 extrapolation in their highest redshift bin.

\subsection{The Build-up of the Stellar Mass}

 The primary goal of measuring the stellar mass density is to
 determine how rapidly the universe assembled its stars.  At
 $z\sim2-3$, our results indicate that the universe only contained
 $\sim 10\%$ of the current stellar mass, regardless of whether we
 refer only to galaxies at \l{V}$> 1.4 \times 10^{10}$\lumsol or
 whether we use the total mass estimates of other authors.  The galaxy
 population in the HDF-S was rich and diverse at $z>2$, but even so it
 was far from finished in its build-up of stellar mass.  By $z\sim 1$,
 however, the total mass density had increased to roughly half its
 local value, indicating that the epoch of $1<z<2$ was an important
 period in the stellar mass build-up of the universe.

 A successful model of galaxy formation must not only explain our
 global results, but also reconcile them with the observed properties
 of individual galaxies at all redshifts.  For example, a population
 of galaxies at $z\sim 1-1.5$ has been discovered (the so called
 extremely red objects or EROs), roughly half of which can be fit with
 formation redshifts higher than 2.4 \citep{Cimatti02} and nearly
 passive stellar evolution thereafter.  Our results, which show that
 the universe contained only $\sim 10\%$ as many stars at $z\sim 2-3$
 as today would seem to indicate that any population of galaxies which
 formed most of its mass at $z\gtrsim2$ can {\it at most} contribute
 $\sim 10\%$ of the present day stellar mass density.  At $z\sim
 1-1.5$, where the EROs reside, the universe had assembled roughly
 half of its current stars.  Therefore, this would imply that the old
 EROs contribute about $\sim 20\%$ of the mass budget at their epoch.
 Likewise, it should be true that a large fraction of the stellar mass
 at low redshift should reside in objects with mass weighted stellar
 ages corresponding to a formation redshift of $1<z<2$.  In support of
 this, \citet{Hogg02} recently have shown that $\sim 40\%$ of the
 local luminosity density at $0.7\mu m$, and perhaps $\sim 50\%$ of
 the stellar mass comes from centrally concentrated, high surface
 brightness galaxies which have red colors.  In agreement with the
 \citet{Hogg02} results, \citet{Bell03a} and \citet{Kauff03} also
 found that $\sim 50-75\%$ of the local stellar mass density resides
 in early type galaxies.  \citet{Hogg02} suggest that their red
 galaxies would have been formed at $z\gtrsim 1$, fully consistent
 with our results for the rapid mass growth of the universe during
 this period.

\section{Summary \& Conclusions}
\label{summary}

 In this paper we presented the globally averaged rest-frame optical
 properties of a \Ks-band selected sample of galaxies with $z<3.2$ in
 the HDF-S.  Using our very deep $0.3-2.2\mu m$, seven band photometry
 taken as part of the FIRE Survey we estimated accurate photometric
 redshifts and rest-frame luminosities for all galaxies with
 \Ksabtot$<25$ and used these luminosity estimates to measure the
 rest-frame optical luminosity density \jrest, the globally averaged
 rest-frame optical color, and the stellar mass density for all
 galaxies at $z<3.2$ with \l{V}$> 1.4\times 10^{10}$\lumsol.  By
 selecting galaxies in the rest-frame \mv-band, we selected them in a
 way much less biased by star formation and dust than the traditional
 selection in the rest-frame UV and much closer to a selection by
 stellar mass.

 We have shown that \jrest in all three bands rises out to $z\sim3$ by
 factors of $4.9\pm1.0$, $2.9\pm0.6$, and $1.9\pm0.4$ in the \u, \mb,
 and \mv-bands respectively.  Modeling this increase in \jrest as an
 increase in \lstar of the local luminosity function, we derive that
 \lstar must have brightened by a factor of 1.7 in the rest-frame
 \mv-band.

 Using our \jrest estimates we calculate the \ubr and \bvr colors of
 all the visible stars in galaxies with \l{V}$>1.4 \times
 10^{10}$\lumsol.  Using the COMBO-17 data we have shown that the mean
 color is much less sensitive to density fluctuations and
 field-to-field variations than either \jrest or \rhostar.  Because of
 their stability, integrated color measurements are ideal for
 constraining galaxy evolution models.  The luminosity weighted mean
 colors lie close to the locus of morphologically normal local galaxy
 colors defined by \citet{Lar78}.  The mean colors monotonically bluen
 with increasing redshift by 0.55 and 0.46 magnitudes in \ubr and \bvr
 respectively out to $z\sim3$.  We interpret this color change
 primarily as a change in the mean stellar age.  The joint colors can
 be roughly matched by simple SFH models if modest amounts of
 reddening ($E(B-V)<0.35$) are applied.  In detail, the redshift
 dependence of \ubr and \bvr cannot be matched exactly by the simple
 models, assuming a constant reddening and constant metallicity.
 However, we show that the models can still be used, even in the face
 of these small disagreements, to robustly predict the stellar
 mass-to-light ratios \mlstarlam{V} of the integrated cosmic stellar
 population implied by our mean rest-frame colors.  Variations in the
 metallicity does not strongly affect this relation and it holds for a
 variety of smooth SFHs.  Even the IMF only affects the normalization
 of this relation, not its slope, assuming that the IMF everywhere is
 the same.  The reddening, which moves objects roughly along this
 relation is, however, a large source of uncertainty.  Using these
 \mlstarlam{V} estimates coupled with our \jrest measurements, we
 derive the stellar mass density \rhostar.  These globally averaged
 estimates of the mass density are more reliable than those obtained
 from the mean of individual galaxies determined using smooth SFHs,
 primarily because the cosmic mean SFH is plausibly much better
 approximated as being smooth, whereas the SFHs of individual galaxies
 are almost definitely not.

 The stellar mass density, \rhostar, increases monotonically with
 increasing cosmic time to come into good agreement with the other
 measured values at $z\lesssim 1$ with a factor of $\sim 10$ increase
 from $z\sim3$ to the present day.  Within the random uncertainties,
 our results agree well with those of \citet{Dick03} in the HDF-N
 although our \rhostar estimates are systematically higher than in the
 HDF-N.  Taken together, the HDF-N and HDF-S paint a picture in which
 only $\sim 5-15\%$ of the present day stellar mass was formed by
 $z\sim 2$.  By $z\sim 1$, however, the stellar mass density had
 increased to $\sim 50\%$ of its present value, implying that a large
 fraction of the stellar mass in the universe today was assembled at
 $1<z<2$.  Our \rhostar estimates slightly underpredict the stellar
 mass density derived from the integral of the SFR($z$) curve at
 $z>2$.  A resolution of the small apparent discrepancy between
 different fields, and between the predictions from optical
 observations will in part require deeper NIR data, to probe further
 down the mass function, and wider fields with multiple pointings to
 control the effects of cosmic variance.  In addition, large amounts
 of follow-up optical/NIR spectroscopy are required to help control
 systematic effects in the \zp estimates.  The 25 square arcminute
 MS1054-03 data taken as part of FIRES and the ACS/ISAAC GOODS
 observations of the CDF-S region will be very helpful for such
 studies.  Observations with SIRTF will also improve the situation by
 accessing the rest-frame NIR, where obscuration by dust becomes much
 less important.  Finally, systematics in the \mlstar estimates may
 exist because of a lack of constraint on the faint end slope of the
 stellar IMF.

 We still have to reconcile global measurements of the galaxy
 population with what we know about the ages and SFHs of individual
 galaxies.  Our globally determined quantities are quite stable and
 may serve as robust constraints on theoretical models, which must
 correctly model the global build-up of stellar mass in addition to
 matching the detailed properties of the galaxy population.

\acknowledgments

GR would like to thank Jarle Brinchmann and Frank van den Bosch for
useful discussions in the process of writing this paper, Christian
Wolf for providing additional COMBO-17 data products, and Eric Bell
and Marcin Sawicki for giving comments on an earlier version of the
paper.  GR would also like to acknowledge the generous travel support
of the Lorentz center and the Leids Kerkhoven-Bosscha Fonds, and the
financial support of Sonderforschungsbereich 375.

Funding for the creation and distribution of the SDSS Archive has been
provided by the Alfred P. Sloan Foundation, the Participating
Institutions, the National Aeronautics and Space Administration, the
National Science Foundation, the U.S. Department of Energy, the
Japanese Monbukagakusho, and the Max Planck Society. The SDSS Web site
is http://www.sdss.org/.

The SDSS is managed by the Astrophysical Research Consortium (ARC) for
the Participating Institutions. The Participating Institutions are The
University of Chicago, Fermilab, the Institute for Advanced Study, the
Japan Participation Group, The Johns Hopkins University, Los Alamos
National Laboratory, the Max-Planck-Institute for Astronomy (MPIA),
the Max-Planck-Institute for Astrophysics (MPA), New Mexico State
University, University of Pittsburgh, Princeton University, the United
States Naval Observatory, and the University of Washington.

\clearpage
\appendix
\section{Derivation of \zp Uncertainty}
\label{app_z_err}

 Given a set of formal flux errors, one way to broaden the redshift
 confidence interval without degrading the accuracy (as noticed in
 R01) is to lower the absolute \chisq of every \chisq$(z)$ curve
 without changing its shape (or the location of the minimum).  By
 scaling up all the flux errors by a constant factor, we can retain
 the relative weights of the points in the \chisq without changing the
 best fit redshift and SED, but we do enlarge the redshift interval
 over which the templates can satisfactorily fit the flux points.
 Since we believe the disagreement between \zs and \zp is due to our
 finite and incomplete template set, this factor should reflect the
 degree of template mismatch in our sample, i.e., the degree by which
 our models fail to fit the flux points.  To estimate this factor we
 first compute the fractional difference between the model and the
 data $\Delta_{i,j}$ for the j$^{\rm th}$ galaxy in the i$^{\rm th}$
 filter,
\begin{equation}
\Delta_{i,j} = \frac{(f^{\rm mod}_{i,j} - f^{\rm dat}_{i,j})}{f^{\rm dat}_{i,j}}
\end{equation}
 where $f^{\rm mod}$ are the predicted fluxes of the best-fit template
 combination and $f^{\rm dat}$ are our actual data.  For each galaxy
 we calculated
\begin{equation}
\Delta_j = \sqrt{\frac{1}{N_{\rm filt} - 1}\sum_{i=2}^{N_{\rm filt}}\Delta_{i,j}^2}
\end{equation}
 where we have ignored the contributions of the \u-band.  While the
 \u-band is important in finding breaks in the SEDs, the exact shapes
 of the templates are poorly constrained blueward of the rest-frame
 \u-band and the \u-data often deviates significantly from the
 best-fit model fluxes.

 To determine the mean deviation of all of the flux points from the
 model $\Delta_{\rm dev}$ we then averaged over all galaxies in our
 complete FIRES sample with \Ksabtot$<22$ (for which the systematic
 \zp errors should dominate over those resulting from photometric
 errors) to obtain
\begin{equation}
\Delta_{dev} = \frac{1}{N_{\rm gal}}\sum_{j=1}^{N_{\rm gal}}\left|\Delta_{j}\right|.
\end{equation}
 We find $\Delta_{\rm dev}= 0.08$, which includes both random and
 systematic deviations from the model.  We modified the Monte-Carlo
 simulation of R01 by calculating, for each object j,
\begin{equation}
\left\langle{\rm \frac{S}{N}}\right\rangle _{j} = \sqrt{\frac{\sum_{i=2}^{N_{\rm filt}}\left(\frac{f_i}{\delta f_i}\right)^2}{N_{\rm filt} - 1}}
\end{equation}
 again excluding the \u-band.  We then scaled the flux errors, for
 each object, using the following criteria:
\begin{equation}
\displaystyle\delta f_i^\prime = \left\{ \begin{array}
    {r@{\quad:\quad}l} \displaystyle\delta f_i &
    \displaystyle\left\langle{\rm \frac{S}{N}}\right\rangle _{j} \leq
    \frac{1}{\Delta_{\rm dev}}\\ \displaystyle\delta f_i \Delta_{\rm
    dev}\left\langle{\rm \frac{S}{N}}\right\rangle _{j} &
    \displaystyle\left\langle{\rm \frac{S}{N}}\right\rangle _{j} >
    \frac{1}{\Delta_{\rm dev}} \end{array} \right ..
\end{equation}
 The photometric redshift error probability distribution is computed
 using the $\delta f_i^\prime$'s.  Note that this procedure will not
 modify the \zp errors of the objects with low $S/N$ where the \zp
 errors are dominated by the formal photometric errors.  The resulting
 probability distribution is highly non-Gaussian and using it we
 calculate the upper and lower 68\% confidence limits on the redshift
 $z_{phot}^{hi}$ and $z_{phot}^{low}$ respectively.  As a single
 number which encodes the total range of acceptable \zp's, we define
 \dzp$\equiv 0.5 * ( z_{phot}^{hi} - z_{phot}^{low})$.

 Figure~6 from L03 shows the comparison of \zp to \zs.  For these
 bright galaxies, it is remarkable that our new photometric redshift
 errorbars come so close to predicting the difference between \zp and
 the true value.  Some galaxies have large \dzp values even when the
 local \chisq minimum is well defined because there is another \chisq
 minimum of comparable depth that is contained in the 68\% redshift
 confidence limits.  There are galaxies with \dzp$<0.05$.  Some of
 these are bright low redshift galaxies with large rest-frame optical
 breaks, which presumably place a strong constraint on the allowed
 redshift.  Many of these galaxies, however, are faint and the \dzp is
 unrealistically low.  Even though these faint galaxies have
 $\left\langle{\rm \frac{S}{N}}\right\rangle _{j} \leq
 \frac{1}{\Delta_{\rm dev}}$, they still can have high $S/N$ in the
 \magb or \magv bandpasses and hence have steep \chisq curves and
 small inferred redshift uncertainties.  In addition, many of these
 galaxies have \zp$>2$ and very blue continuum longward of \Lya.  The
 imposed sharp discontinuity in the template SEDs at the onset of HI
 absorption causes a very narrow minimum in the \chisq$(z)$ curve, and
 hence a small \dzp, but likely differs from the true shape of the
 discontinuity because we use the mean opacity values of
 \citet{Mad95}, neglecting its variance among different lines of
 sight.

 It is difficult to develop a scheme for measuring realistic
 photometric redshift uncertainties over all regimes.  The \dzp
 estimate derives the \zp uncertainties individually for each object,
 but can underpredict the uncertainties in some cases.  Compared to
 the technique of R01 however, a method based completely on the
 Monte-Carlo technique is preferable because it has a
 straightforwardly computed redshift probability function.  This trait
 is desirable for estimating the errors in the rest-frame luminosities
 and colors and for this reason we will use \dzp as our uncertainty
 estimate in this paper.

\section{Rest-Frame Photometric System}
\label{photsys}

To define the rest-frame \u, \mb, and \mv fluxes we use the filter
transmission curves and zeropoints tabulated in \citet{Bess90},
specifically his $UX$, $B$, and $V$ filters.  The Bessell zeropoints
are given as magnitude offsets with respect to a source which has
constant $f_{\nu}$ and $AB=0$.  The $AB$ magnitude is defined as
\begin{equation}
  AB_{\nu} = -2.5 * {\rm log}_{10} \langle f_{\nu}\rangle - 48.58
\end{equation}
where $\langle f_{\nu}\rangle$ is the flux $f_{\nu}(\nu)$ observed
through a filter $T(\nu)$ and in units of $ergs~s^{-1} cm^{-2}
Hz^{-1}$.  Given the zeropoint offset $ZP_{\nu}$ for a given filter,
the Vega magnitude $m_{\nu}$ is then
\begin{equation}
  m_{\nu} = AB_{\nu} - ZP_{\nu} = -2.5 * {\rm log}_{10} \langle f_{\nu}\rangle - 48.58 - ZP_{\nu}.
  \label{mvegaeq}
\end{equation}
All of our observed fluxes and rest-frame template fluxes are
expressed in $f_{\lambda}$.  To obtain rest-frame magnitudes in the
\citet{Bess90} system, we must calculate the conversion from
$f_{\lambda}$ to $f_{\nu}$ for the redshifted rest-frame filter set.
The flux density of an SED with $f_{\lambda}(\lambda)$ integrated
through a given filter with transmission curve $T(\lambda)$ is
\begin{equation}
  \langle f_{\lambda} \rangle = \frac{\int f_{\lambda}(\lambda') T'(\lambda') d\lambda'}{\int T'(\lambda') d\lambda'}
\end{equation}
or
\begin{equation}
  \langle f_{\nu} \rangle = \frac{\int f_{\nu}(\nu') T'(\nu') d\nu'}{\int T'(\nu') d\nu'}.
\end{equation}
Since 
\begin{equation}
  \int f_{\lambda}(\lambda') T'(\lambda') d\lambda' = \int f_{\nu}(\nu') T'(\nu') d\nu'
\end{equation}
we can convert to $\langle f_{\nu} \rangle$ through
\begin{equation}
  \langle f_{\nu} \rangle = \langle f_{\lambda} \rangle * \frac{\int T'(\lambda') d\lambda'}{\int T'(\nu') d\nu'}
\end{equation}
and use $\langle f_{\nu} \rangle$ to calculate the apparent rest-frame
Vega magnitude through the redshifted filter via Eq.~\ref{mvegaeq}.

\section{Estimating Rest-Frame Luminosities}
\label{lumder}

 We derive for any given redshift, the relation between the apparent
 AB magnitude $m_{\lambda_{z}}$ of a galaxy through a redshifted
 rest-frame filter, its observed fluxes $\langle f_{\lambda_i,obs}
 \rangle$ in the different filters $i$, and the colors of the spectral
 templates.  At redshift z, the rest-frame filter with effective
 wavelength $\lambda_{rest}$ has been shifted to an observed
 wavelength
\begin{equation}
  \lambda_z = \lambda_{rest} \times ( 1 + z) 
\label{redeq}
\end{equation}
and we define the adjacent observed bandpasses with effective
wavelengths $\lambda_l$ and $\lambda_h$ which satisfy
\begin{equation}
\lambda_l<\lambda_z\leq\lambda_h.
\label{lameq}
\end{equation}
We now define
\begin{equation}
  C_{obs} \equiv m_{obs,\lambda_l} - m_{obs,\lambda_h}
\end{equation}
where $m_{obs,\lambda_l}$ and $m_{obs,\lambda_h}$ are the AB
magnitudes which correspond to the fluxes $\langle f_{\lambda_l,obs}
\rangle$ and $\langle f_{\lambda_h,obs} \rangle$ respectively.  We
then shift each template in wavelength to the redshift z  and compute,
\begin{equation}
  C_{templ} \equiv m_{templ,\lambda_l} - m_{templ,\lambda_h},
\end{equation}
where $m_{templ,\lambda_l}$ and $m_{templ,\lambda_h}$ are the AB
magnitudes through the $\lambda_l$ and $\lambda_h$ observed bandpasses
(including the atmospheric and instrument throughputs).  We sort the
templates by their $C_{templ}$ values, $C_{templ,a}$, $C_{templ,b}$,
etc., and find the two templates such that
\begin{equation}
  C_{templ,a} \leq C_{obs} < C_{templ,b}.
\end{equation}
We then define for the $a^{\rm th}$ template 
\begin{equation}
  C_{\lambda_l,z,a} \equiv m_{templ,\lambda_l} - m_{templ,\lambda_{z}}
\end{equation}
where $m_{templ,\lambda_{z}}$ is the apparent AB magnitude of the
redshifted $a^{\rm th}$ template through the redshifted
$\lambda_{rest}$ filter.  We point out that because our computations
always involve colors, they are not dependent on the actual template
normalization (which cancels out in the difference).  Taking our
observed color $C_{obs}$ and the templates with adjacent ``observed''
colors $C_{templ,a}$ and $C_{templ,b}$, we can interpolate between
$C_{\lambda_l,z,a}$ and $C_{\lambda_l,z,b}$
\begin{equation}
  m_{obs,\lambda_l} - m_{\lambda_{z}} = C_{\lambda_l,z,a} + (C_{obs} - C_{templ,a}) \times \left(\frac{C_{\lambda_l,z,b} - C_{\lambda_l,z,a}}{C_{templ,b} - C_{templ,a}}\right)
\label{restmageq}
\end{equation}
and solve for $m_{\lambda_{z}}$.

 When $C_{obs}$ lies outside the range of the $C_{templ}$'s, we simply
 take the two nearest templates in observed $C_{templ}$ space and
 extrapolate Eq.~\ref{restmageq} to compute $m_{\lambda_{z}}$.

 Equation~\ref{restmageq} has the feature that $m_{\lambda_{z}}
 \approx m_{obs,\lambda_l}$ when $\lambda_z = \lambda_l$ (and hence
 when $C_{\lambda_l,z,a}$ and $C_{\lambda_l,z,b} \approx 0$).  While
 this method still assumes that the templates are reasonably good
 approximations to the true shape of the SEDs it has the advantage
 that it does not rely on exact agreement.  Galaxies whose observed
 colors fall outside the range of the templates can also be easily
 flagged.  A final advantage of this method is that the uncertainty in
 $m_{\lambda_{z}}$ can be readily calculated from the errors in the
 observed fluxes.

 From $m_{\lambda_{z}}$, we compute the rest-frame luminosity by
 applying the K-correction and converting to luminosity units
\begin{equation}
  \frac{\rm L^{rest}}{\rm L_{\odot}} = 10^{\displaystyle -0.4(m_{\lambda_{z}} - {\rm M_{\odot,\lambda_{rest}}} - ZP_{\lambda_{rest}})}\times \left(\frac{\rm D_L}{10pc}\right)^2 \times (1 + z)^{-1} \times h^{-2}
\end{equation}
 where ${\rm M_{\odot,\lambda_{rest}}}$ is the absolute magnitude of
 the sun in the $\lambda_{rest}$ filter (${\rm M_{\odot,U}=+5.66}$,
 ${\rm M_{\odot,B}=+5.47}$, and ${\rm M_{\odot,V}=+4.82}$ in Vega
 magnitudes; Cox 2000), $ZP_{\lambda_{rest}}$ is the zeropoint in that
 filter (as in Eq.  \ref{mvegaeq} but expressed at $\lambda$ and not
 at $\nu$), and ${\rm D_L}$ is the distance modulus in parsecs.
 Following R01, we correct this luminosity by the ratio of the
 $K_s^{\rm tot}$ flux to the modified isophotal aperture flux (see
 L03).  This adjustment factor, which accounts both for the larger
 size of the total aperture and the aperture correction, changes with
 apparent magnitude and it ranges from 1.23, at $20<$\Ksabtot$\leq24$,
 to 1.69, at $24<$\Ksabtot$\leq25$, and it has an RMS dispersions of
 0.17 and 0.49 in the two magnitude bins respectively.

 The uncertainty in the derived \lrest has contributions both from the
 observational flux errors and from the redshift uncertainty, which
 causes $\lambda_z$ to move with respect to the observed filters.  The
 first effect is estimated by propagating the observed flux errors
 through Eq.~\ref{restmageq}.  As an example, object 531 at \zp=2.20
 has \Ksabtot$=24.91$ and signal-to-noise in the \Ks-band of 8.99 and
 5.43 in our modified isophotal and total apertures respectively.  The
 resultant error in \l{V} purely from flux errors is then $26\%$.  At
 \Ksabtot~$\approx 24$, the typical signal-to-noise in the \Ks-band
 increases to $\approx 13$ and $\approx 6.3$ in our modified isophotal
 and total apertures respectively and the error \lrest decreases
 accordingly.

 To account for the redshift dependent error in the calculated
 luminosity, we use the Monte-Carlo simulation described first in R01
 and updated in \S\ref{app_z_err}.  For each Monte-Carlo iteration we
 calculate the rest-frame luminosities and determine the 68\%
 confidence limits of the resulting distribution.  The 68\% confidence
 limits in \lrest can be highly asymmetric, just as for \zp.  For
 objects with \Ksabtot~$\lesssim 25$ we find that the contributions to
 the total \lrest error budget are dominated by the redshift errors
 rather than by the flux errors.

\end{document}